\documentclass[pra,twocolumn,groupedaddress,showpacs,floatfix,accepted=2017-12-15]{quantumarticle}

\usepackage{graphics}
\usepackage{graphicx}
\usepackage{bm}
\usepackage{amsmath}
\usepackage{amsfonts} 
\usepackage{amssymb}
\usepackage{latexsym}
\usepackage{color}
\usepackage{mathrsfs}
\usepackage{braket}
\usepackage{enumerate}
\usepackage{hyperref}
\usepackage{float}
\usepackage{mathtools}
\usepackage[raggedright,tight]{subfigure}  

\begin{document}

\title{Vacuum source-field correlations and advanced waves in quantum optics}
\author{Adam Stokes}

\affiliation{Photon Science Institute, University of Manchester, Oxford Road, Manchester, M13 9PL, United Kingdom}

\begin{abstract}
The solution to the wave equation as a Cauchy problem with prescribed fields at an initial time $t=0$ is purely retarded. Similarly, in the quantum theory of radiation the specification of Heisenberg picture photon annihilation and creation operators at time $t>0$ in terms of operators at $t=0$ automatically yields purely retarded source-fields. However, we show that two-time quantum correlations between the retarded source-fields of a stationary dipole and the quantum vacuum-field possess advanced wave-like contributions. Despite their advanced nature, these correlations are perfectly consistent with Einstein causality. It is shown that while they do not significantly contribute to photo-detection amplitudes in the vacuum state, they do effect the statistics of measurements involving the radiative force experienced by a point charge in the field of the dipole. Specifically, the dispersion in the charge's momentum is found to increase with time. This entails the possibility of obtaining direct experimental evidence for the existence of advanced waves in physical reality, and provides yet another signature of the quantum nature of the vacuum.
\end{abstract}

\maketitle
 
\section{Introduction}\label{int}

Electrodynamics describes the production, propagation and absorption of electromagnetic waves. The problem most commonly encountered in classical electrodynamics is that in which the source is specified at some initial time, and then the electromagnetic fields are sought at later times. The unique solution to this problem is fully retarded, which means that the electromagnetic source-fields depend in a causal way on the state of the source in the past \cite{jackson_classical_1998,barton_elements_1989,zangwill_modern_2012}. Advanced source-fields, which depend on the source in the future are obtained when Maxwell's equations are solved running into the past \cite{barton_elements_1989,spohn_dynamics_2007}. These solutions are not viewed as physical within the context of an initial value problem.

It is well known that in quantum optics the commutators of the free electromagnetic fields involve both retarded and advanced green's functions for the wave-operator \cite{cohen-tannoudji_photons_1997}. On the other hand the subject of advanced waves in the presence of sources has received more limited attention in quantum theory. The reason for this may be that the problem of finding the Heisenberg picture source-fields at time $t>0$ in terms of Schr\"odinger picture operators at time $t=0$ is an initial value problem whose solution entails purely retarded source-fields. However, due to differences between quantum and classical theories the complete absence of advanced contributions in quantum optics is not immediate. The fluctuating quantum vacuum offers one of the most striking examples of such a difference. The theories also differ in their treatment of electromagnetic correlations. 

The quantum vacuum is used to interpret various phenomena in quantum optics including spontaneous emission \cite{milonni_interpretation_1973}, atomic energy-shifts \cite{power_zero-point_1966,welton_observable_1948}, and Casimir forces \cite{casimir_influence_1948}. Together with the laws of space and time the quantum theory of the vacuum predicts intriguing physical phenomena such as Hawking radiation associated with black holes \cite{hawking_particle_1975} and the Unruh-Davies effect whereby an accelerated detector in the vacuum of Minkowski spacetime registers thermal excitations \cite{unruh_notes_1976,davies_scalar_1975}. The Unruh-Davies effect and related effects \cite{olson_extraction_2012} can be understood in terms of correlations between the quantum vacuum at timelike separated points in space-time. Vacuum induced correlations across spacelike separations have also received attention \cite{reznik_violating_2005}.

In atomic and optical physics many effects attributed to the quantum vacuum can often also be understood, at least partially, in terms of radiation reaction \cite{milonni_quantum_1994,dalibard_vacuum_1982,jaffe_casimir_2005}. Thus, identifying specific signatures associated with the quantum vacuum remains of broad interest in quantum optics. Recently, direct observation of vacuum fluctuations has been achieved \cite{riek_direct_2015,moskalenko_paraxial_2015,riek_subcycle_2017}, which opens up new possibilities for investigating the interplay between vacuum fluctuations and the electromagnetic fields produced by charged sources.

In this paper we consider the fields produced by a quantum dipole. We show that the presence of the quantum vacuum and the distinct nature of quantum correlations imply that certain physical predictions involve advanced waves. More specifically, we show that the combination of purely retarded source-fields and quantum vacuum fields result in advanced-wave contributions to two-time quantum correlations.

There are seven sections to this paper. We begin in section \ref{clas} by briefly reviewing the interpretation of waves in classical physics. In section \ref{bckgrnd} we introduce a model of the dipole-field system and derive the electromagnetic source-fields of the dipole. In section \ref{vscri} we discern the role of interference between vacuum and source-fields within the process of spontaneous emission. In section \ref{vscaw} we consider general two-time quantum correlations functions of the electromagnetic field, and show that vacuum-source correlations receive advanced-wave contributions. Our subsequent aim is to determine whether such correlations might be physically measurable. In section \ref{pd} we briefly verify that vacuum-source correlations do not significantly contribute to photo-detection amplitudes. In section \ref{forces} we consider the radiative force experienced by a test charge in the field of a quantum dipole. By considering a charge with relatively slow dynamics, we are able to develop a quantum description of the free charge, which is analogous to the multipolar description of bound charge systems. This description enables us to properly identify the relevant vacuum and source-fields. We then show that vacuum-source correlations make signficant contributions to the statistical predictions involving the force experienced by the test charge. Finally in section \ref{conc} we summarise our findings.

\section{Classical solutions of the wave equation}\label{clas}

We begin by briefly reviewing the interpretation of solutions to the wave equation in classical physics. This interpretation may be contrasted with the interpretation given to the advanced-wave like contributions to quantum correlations that are reported in subsequent sections.

The inhomogeneous wave equation for a scalar field $\psi$ over Minkowski spacetime is $\left[\partial_t^2-\nabla^2 \right] \psi(t,{\bf x}) = f(t,{\bf x})$, where we have assumed units such that the speed of the wave is one and where $f(t,{\bf x})$ describes the sources of the waves. A well-posed problem is one in which boundary conditions and the source data $f$ are specified, along with the fields $\psi(t_0,{\bf x})$ and $\psi_{t_0}(t_0,{\bf x}) =\partial_{t_0}\psi(t_0,{\bf x})$ at some time $t_0$ \cite{barton_elements_1989}. The fields at some other time $t$ are then sought.

Methods of solving the wave equation in classical electrodynamics invariably involve green's functions for the wave-operator $\partial_t^2-\nabla^2$ \cite{jackson_classical_1998,zangwill_modern_2012,spohn_dynamics_2007}. The retarded and advanced green's functions $G^+$ and $G^-$ admit the representations
\begin{align}
G^{\pm}(t,{\bf x}|t',{\bf x}') = {1\over 4\pi |{\bf x}-{\bf x}'|}\delta\left(t'-t\pm |{\bf x}-{\bf x}'|\right).
\end{align}
In conjunction with either initial or final conditions the wave equation is easily solved in reciprocal space \cite{spohn_dynamics_2007}. This method of solution is closest to that used in quantum theory. The quantum electric field associated with a point source within the Coulomb gauge is found using essentially this method in appendix \ref{ap4}.

If initial conditions are specified such that before any sources are activated at $t=0$ there is a known input configuration $\psi_{\rm in}$, which satisfies the homogenous equation $\left[\partial_t^2-\nabla^2 \right] \psi_{\rm in}(t,{\bf x})=0$, the field at time $t>0$ is given by \cite{spohn_dynamics_2007}
\begin{align}\label{solr2}
\psi(t,{\bf x}) =& \psi_{\rm in}(t,{\bf x}) \nonumber \\ &+ \int d^3x' \int_0^t dt' f(t',{\bf x})G^+(t,{\bf x}|t',{\bf x}').
\end{align}
Significantly, the advanced green's function $G^-$ does not contribute to the solution for $t>0$ of the initial value problem. Moreover, when initial conditions are specified for the source function $f$, the homogeneous component $\psi_{\rm in}$ can be taken to vanish \cite{zangwill_modern_2012}. Therefore, in classical electrodynamics only the {\em retarded source-field} plays a role in the physical {\em predictions} associated with a source that is initially fully specified. The retarded solution is replaced by the advanced solution when final conditions are specified instead of initial conditions, and the Maxwell equations are solved running into the past \cite{barton_elements_1989,jackson_classical_1998,spohn_dynamics_2007}.

In contrast to the classical situation, in quantum theory, even when all sources are fully specified the vacuum plays an important role in physical phenomena \cite{milonni_quantum_1994}. We will show further that although all source-fields remain fully retarded, certain predictions feature advanced waves associated with the quantum vacuum.

\section{Dipole-field Hamiltonian and electric and magnetic source-fields}\label{bckgrnd}

\subsection{Dipole-field Hamiltonian}\label{hamf}

Consider a single atom coupled to the electromagnetic field. Within the electric dipole approximation the Hamiltonian describing this system is $H=H_0+V$ with
\begin{align}\label{h}
H_0 &=\sum_n \omega_n\ket{n}\bra{n} + \int d^3 k \sum_\lambda \omega \left( a^\dagger_\lambda({\bf k}) a_\lambda({\bf k}) +{1\over 2} \right)\nonumber \\  &=H_d+H_f \nonumber \\ V&=- {\hat {\bf d}}\cdot {\bf D}_{\rm T}({\bf 0})
\end{align}
where ${\ket n}$ denotes an energy eigenstate of the dipole with associated energy $\omega_n$, $a_\lambda({\bf k})$ denotes the annihilation operator for a photon with momentum ${\bf k}$ and polarisation $\lambda$, ${\hat {\bf d}}$ denotes the dipole moment operator, and ${\bf D}_{\rm T}$ denotes the transverse electric displacement field. The dipole is assumed to be located at the origin ${\bf 0}$.

In terms of photonic operators $a_\lambda({\bf k})$ the transverse displacement field admits the mode expansion
\begin{align}\label{D}
&{\bf D}_{\rm T}(t,{\bf x}) ={\bf D}_{\rm T}^{(+)}(t,{\bf x})+{\bf D}_{\rm T}^{(-)}(t,{\bf x}),\nonumber \\
&{\bf D}_{\rm T}^{(+)}(t,{\bf x}) = i\int d^3 k \sum_\lambda \sqrt{\omega \over 2(2\pi)^3} {\bf e}_\lambda({\bf k}) a_\lambda(t,{\bf k})e^{i{\bf k}\cdot {\bf x}} \nonumber \\ &\qquad \qquad  ~={\bf D}_{\rm T}^{(-)}(t,{\bf x})^\dagger
\end{align}
where the ${\bf e}_{1,2}({\bf k})$ are mutually orthogonal unit polarisation vectors orthogonal to ${\bf k}$, and $\omega=|{\bf k}|$. The vacuum (free) components of the electric field and transverse displacement field are identical for all ${\bf x}$ and are given by
\begin{align}\label{Dfree}
{\bf D}_{{\rm T},0}^{(+)}(&t,{\bf x}) ={\bf E}_0^{(+)}(t,{\bf x}),\nonumber \\
&=i\int d^3 k \sum_\lambda \sqrt{\omega \over 2(2\pi)^3} {\bf e}_\lambda({\bf k}) a_\lambda({\bf k})e^{-i\omega t + i{\bf k}\cdot {\bf x}}.
\end{align}
%

For ${\bf x}\neq {\bf 0}$ the source components also coincide ${\bf D}_{{\rm T},s}={\bf E}_s$. In terms of photonic operators the electric field for ${\bf x}\neq {\bf 0}$ is therefore given by
\begin{align}\label{E}
&{\bf D}_{\rm T}(t,{\bf x}) = {\bf E}(t,{\bf x}) = {\bf E}^{(+)}(t,{\bf x})+{\bf E}^{(-)}(t,{\bf x}) \nonumber \\
& {\bf E}^{(+)}(t,{\bf x}) ={\bf D}_{\rm T}^{(+)}(t,{\bf x}).
\end{align}
The situation is different at the position of the dipole ${\bf x} ={\bf 0}$. This is most easily seen by noting that ${\bf D}({\bf x}) - {\bf E}({\bf x}) = {\bf P}({\bf x})$ where ${\bf P}({\bf x}) = {\bf d}\delta({\bf x})$ in the electric dipole approximation. Thus, at the position of the dipole ${\bf E}$ differs from ${\bf D}$ by an infinite field. Since radiation is detected by an absorbing atom at the position of the atom, it is important to recognise the distinction between the electric field and the displacement field at this position. The magnetic field admits the mode-expansion
\begin{align}
&{\bf B}(t,{\bf x}) ={\bf B}^{(+)}(t,{\bf x})+{\bf B}^{(-)}(t,{\bf x}),\nonumber \\ 
&{\bf B}^{(+)}(t,{\bf x})\nonumber \\ &= i\int d^3 k \sum_\lambda \sqrt{\omega \over 2(2\pi)^3} {\hat {\bf k}}\times {\bf e}_\lambda({\bf k}) \, a_\lambda(t,{\bf k})e^{i{\bf k}\cdot {\bf x}}
\end{align}
and similarly to the transverse displacement field it can be split into vacuum and source components via the corresponding partitioning of $a_\lambda(t,{\bf k})$.

\subsection{Electric and magnetic source-fields}\label{eandbfs}

The equation of motion for the photon annihilation operator is found using the Hamiltonian (\ref{h}) and once formally integrated reads
\begin{align}\label{a2}
a_\lambda(t,{\bf k})  =&\,  a_\lambda({\bf k})e^{-i\omega t} \nonumber \\ &+ \int_0^t dt' \, e^{-i\omega(t-t')}\sqrt{\omega\over 2(2\pi)^3}{\bf e}_\lambda({\bf k})\cdot{\hat {\bf d}}(t').
\end{align}
Away from the dipole, the electric and magnetic dipole source-fields are found using Eqs.~(\ref{E}) and (\ref{a}) and are exactly analogous to classical fields of an oscillating dipole;
\begin{align}\label{Erad}
{\bf E}_s(t,{\bf x}) =&{1\over 4\pi x^3}\left\{3{\hat {\bf x}}\left[{\hat {\bf x}}\cdot {\hat {\bf d}}(t_r)\right]-{\hat {\bf d}}(t_r)\right\}\nonumber \\ &+{1\over 4\pi x^2}\left\{3{\hat {\bf x}}\left[{\hat {\bf x}}\cdot {\dot {\hat {\bf d}}}(t_r)\right]-{\dot {\hat {\bf d}}}(t_r)\right\} \nonumber \\ &+{1\over 4\pi x} \left\{{\hat {\bf x}}\left[{\hat {\bf x}}\cdot {\ddot {\hat {\bf d}}}(t_r)\right]-{\ddot {\hat{\bf d}}}(t_r)\right\},\nonumber \\ 
{\bf B}_s(t,{\bf x}) =& -{1\over 4\pi x}{\hat {\bf x}}\times {\ddot {\hat {\bf d}}}(t_r) - {1\over 4\pi x^2}{\hat {\bf x}}\times {\dot {\hat {\bf d}}}(t_r)
\end{align}
where $t_r=t-x>0$. For $t<x$ the source-fields vanish.

The source-fields can be split into radiative and non-radiative components. The radiation fields are responsible for irreversible energy loss from the dipole. They vary as $x^{-1}$, are orthogonal to ${\bf x}$ and depend on the dipole moment acceleration. They are related via ${\bf E}_{{\rm rad},s}(t,{\bf x}) = -{\hat {\bf x}} \times {\bf B}_{{\rm rad},s}(t,{\bf x})$. The remaining parts of the source-fields are non-radiative. The non-radiative electric field includes both velocity-dependent and static Coulomb-like parts whereas the non-radiative magnetic field is entirely velocity-dependent.

The creation and annihilation source-fields ${\bf E}^{(\pm)}_s$ are not causal, that is, they do not generally vanish for $t<x$. However, as in references \cite{milonni_photodetection_1995,carmichael_statistical_1999} we can perform both a rotating-wave approximation and an associated Markov approximation, which yield for $t>x$
\begin{align}\label{Epms}
{\bf E}^{(+)}_s&(t,{\bf x}) =\nonumber \\ &\sum_{\substack{n,m \\n<m}}\bigg[\left\{3{\hat {\bf x}}\left({\hat {\bf x}}\cdot {\bf d}_{nm}\right)-{\bf d}_{nm}\right\}\left({i\omega_{nm}\over 4\pi x^2}+{1\over 4\pi x^3}\right) \nonumber \\ &+ {\omega_{nm}^2\over 4\pi x} \left\{{\bf d}_{nm}-{\hat {\bf x}}({\hat {\bf x}}\cdot {\bf d}_{nm})\right\}\bigg]\sigma_{nm}(t_r), \nonumber \\ 
{\bf B}^{(+)}_s&(t,{\bf x})=\sum_{\substack{n,m \\n<m}}{\hat {\bf x}}\times{\bf d}_{nm}\bigg[{\omega_{nm}^2\over 4\pi x}-{i\omega_{nm}\over 4\pi x^2}\bigg]\sigma_{nm}(t_r)
\end{align}
where $\sigma_{nm}(t)= e^{iH t}\ket{n}\bra{m}e^{-iHt}$. For $t<x$ the creation and annihilation fields vanish within the approximations made. These approximations also ensure that the positive and negatve frequency fields are associated with dipole lowering and raising operators respectively. Eq.~ (\ref{Epms}) can be further simplified through the approximation $\sigma_{nm}(t) \approx \sigma_{nm}(0)e^{i\omega_{nm}t}$. The latter is appropriate for perturbative evaluations of quadratic field functionals correct to second order. The steps leading to the final result in Eq.~(\ref{Epms}) are given in appendix \ref{ap0}.

In later sections we we will focus on the case of a two-level dipole with ground and excited states denoted $\ket{g}$ and $\ket{e}$ respectively. The two-level transition frequency is denoted $\omega_0$, and the transition dipole moment, which is assumed to be real, is denoted ${\bf d}$. We also introduce the two-level operators
\begin{align}
\sigma^+ = \ket{e}\bra{g},\qquad \sigma^- =(\sigma^+)^\dagger,\qquad \sigma^z = [\sigma^+,\sigma^-]
\end{align}
in terms of which the source-fields can be written
\begin{align}\label{Ema}
&{\bf E}^{(+)}_s(t,{\bf x}) ={\bm {\mathcal E}}^*({\bf x}) \sigma^-(t_r),\nonumber \\
&{\bf B}^{(+)}_s(t,{\bf x})={\bm {\mathcal B}}^*({\bf x})\sigma^-(t_r)
\end{align}
wherein for convenience we have defined
\begin{align}\label{ebcoeffs}
{\bm {\mathcal E}}({\bf x})=&\left({i\omega_0\over 4\pi x^2}+{1\over 4\pi x^3}\right) \left\{3{\hat {\bf x}}\left({\hat {\bf x}}\cdot {\bf d}\right)-{\bf d}\right\}\nonumber \\ &+ {\omega_0^2\over 4\pi x} \left\{{\bf d}-{\hat {\bf x}}({\hat {\bf x}}\cdot {\bf d})\right\},\nonumber \\ 
{\bm {\mathcal  B}}({\bf x})=&\left({\omega_0^2\over 4\pi x} - {i\omega_0\over 4\pi x^2}\right){\hat {\bf x}}\times{\bf d}.
\end{align}

We remark finally that it is not obvious that the equal-time commutation relations between the fields whose source components are given in Eq.~(\ref{Ema}) are the same as those for the fields in Eq.~(\ref{E}), because approximations have been made. That the commutation relations do remain the same is however necessary in order that the resulting theory is formally consistent. We prove that the equal-time commutators are in fact preserved in appendix \ref{ap3}.

\section{Vacuum-source correlations and radiation intensity}\label{vscri}

Before considering general quadratic functionals (correlation functions) of the electromagnetic fields we consider in detail a particular example - the radiation intensity. Our aim is to identify the contribution made by the quantum vacuum, and in particular, the role played by vacuum source-field interference. The treatment in this section extends previous perturbative treatments found in \cite{power_quantum_1992,salam_molecular_2008,salam_molecular_2009,stokes_quantum_2016}.

\subsection{Radiated power}\label{rp}

In order to identify the radiation intensity, which is the radiated power per unit area, we begin by calculating the radiated power due to the spontaneous photon emission of an initially excited dipole. The $S$-matrix element representing the probability amplitude for the spontaneous emission process $\ket{e,0}\to \ket{m,{\bf k}\lambda}$ with $m<e$ and $\ket{{\bf k}\lambda}$ the state of a single photon, can be calculated easily using second order perturbation theory. The associated rate at which this process occurs is found using Fermi's golden rule to be
\begin{align}\label{spont}
\Gamma_{em} = {\omega_{em}^3|{\bf d}_{em}|^2 \over 3\pi}
\end{align}
where $\omega_{em} = \omega_e-\omega_m$ and ${\bf d}_{em} = \bra{e}{\bf d}\ket{m}$. The $S$-matrix consists of probability amplitudes for processes that conserve the bare energy $H_0$. Thus, in spontaneous emission the energy of the resulting photon is precisely the energy of the atomic transition through which it was emitted. The energy flux in the process $\ket{e,0}\to \ket{m,{\bf k}\lambda},~m<e$ is therefore $\omega_{em}\Gamma_{em}$, which implies that the total radiated power is to second order given by
\begin{align}\label{P}
P = \sum_{m<e} \omega_{em}\Gamma_{em}.
\end{align}
The above derivation of $P$ relies entirely on the calculation of matrix elements of the interaction Hamiltonian, so it does not involve using any electromagnetic fields away from the source at ${\bf 0}$. In fact, since ${\bf D}_{\rm T}({\bf 0}) \neq {\bf E}({\bf 0})$, the electric and magnetic fields do not directly feature anywhere in the above calculation.

\subsection{Derivation using electromagnetic fields}\label{dem}

An alternative expression for the radiated power that does involve the electromagnetic fields, can be obtained from Poynting's theorem. The energy flux radiated by the dipole located at the centre of a sphere with radius $x$ is given by
\begin{align}\label{Poynt}
P =\lim_{x\to \infty} \int d\Omega  \, x^2  I(t,{\bf x}) =  \int d\Omega  \, x^2  I_{\rm rad}(t,{\bf x})
\end{align}
where the integration is performed over all directions. Eq.~ (\ref{Poynt}) is definitive of the {\em radiation intensity} $I_{\rm rad}$, which is the radiated power per unit area. The intensity component $I_{\rm rad}$ is a quadratic functional of the radiation fields, which vary as $x^{-1}$.

We now proceed to determine the radiated power using the fields in section \ref{hamf}. In order to compare the result of Eq.~(\ref{Poynt}) with Eq.~(\ref{P}), which was obtained using second order perturbation theory we make use of Eq.~(\ref{Epms}) coupled with the perturbative approximation $\sigma_{nm}(t) \approx \sigma_{nm}(0)e^{i\omega_{nm}t}$. First let us suppose that the intensity $I_{\rm rad}$ in Eq.~(\ref{Poynt}) is of the Glauber type 
\begin{align}\label{IG}
I_G(t,{\bf x}) &=\langle {\bf E}_{\rm rad}^{(-)}(t,{\bf x})\cdot {\bf E}_{\rm rad}^{(+)}(t,{\bf x}) \rangle_{0,e} \nonumber \\ &=\langle {\bf E}^{(-)}_{{\rm rad},s}(t,{\bf x})\cdot {\bf E}^{(+)}_{{\rm rad},s}(t,{\bf x})\rangle_{e;0}
\end{align}
where the average is taken in the state $\ket{e,0}$, and we have retained only the radiative contribution dependent on ${\bf E}_{\rm rad}={\bf E}_0+{\bf E}_{{\rm rad},s}$. Letting $\sigma_{nm}(t) \approx \sigma_{nm}(0)e^{i\omega_{nm}t}$ in Eq.~(\ref{Epms}) we find that
\begin{align}\label{IG2}
I_G(t,{\bf x}) &=\left({1\over 4\pi x}\right)^2 \sum_{m<e} \omega_{em}^4  [{\bf d}_{em}-{\hat {\bf x}}({\hat {\bf x}}\cdot {\bf d}_{em})]^2.
\end{align}
Using Eq.~(\ref{IG}), the corresponding radiated power is
\begin{align}\label{Ppm}
P_G = \int d\Omega  \, x^2  I_G(t,{\bf x})= {1\over 2} \sum_{m<e} \omega_{em}\Gamma_{em},
\end{align}
which is only half of the power given in Eq.~(\ref{P}). The correct power is evidently obtained from the intensity $2I_G$ rather than $I_G$ itself

\subsection{General definition of the radiation Intensity}\label{gdri}

In order to motivate a general definition of the radiation intensity let us first consider the case of a free field confined to a volume $V$. The vacuum energy $H_{\rm vac}$ and the associated density $H_{\rm vac}/V$ are not measurable. Only differences in energy-density from the vacuum value are measurable. The integral of the measurable local energy-density over all space should therefore yield the energy difference $H_f-H_{\rm vac}$ \cite{sipe_photon_1995,bialynicki-birula_v_1996}.

Quite generally, the vacuum energy density can be expressed as
\begin{align}
U_{\rm vac} &= {H_{\rm vac} \over V} = \langle {\bf E}_0^{(+)}(t,{\bf x}) \cdot {\bf E}_0^{(-)}(t,{\bf x})\rangle_0 \nonumber \\ &=[E_{0,i}^{(+)}(t,{\bf x}), E_{0,i}^{(-)}(t,{\bf x})] \nonumber \\ &= [B_{0,i}^{(+)}(t,{\bf x}),B_{0,i}^{(-)}(t,{\bf x})].
\end{align}
Thus, in the general case in which sources are present the average electromagnetic energy-density can be written
\begin{align}\label{U}
&U_{\rm em}(t,{\bf x})-U_{\rm vac}\equiv \nonumber \\  &{1\over 2} \left[{\bf E}(t,{\bf x})^2+ {\bf B}(t,{\bf x})^2\right] - [E_{0,i}^{(+)}(t,{\bf x}), E_{0,i}^{(-)}(t,{\bf x})].
\end{align}
Since the total fields ${\bf E}^{(\pm)}$ and the vacuum fields ${\bf E}_0^{(\pm)}$ satisfy the same equal-time commutation relations (cf appendix \ref{ap3}),
\begin{align}\label{etc}
&[E_{0,i}^{(+)}(t,{\bf x}), E_{0,j}^{(-)}(t,{\bf x}')] = [E_i^{(+)}(t,{\bf x}),E_j^{(-)}(t,{\bf x}')] \nonumber \\ =\, &[B_i^{(+)}(t,{\bf x}),B_j^{(-)}(t,{\bf x}')]= [B_{0,i}^{(+)}(t,{\bf x}),B_{0,j}^{(-)}(t,{\bf x}')],
\end{align}
for the total energy density $U=U_{em}+U_{\rm matter}$, where $U_{\rm matter}$ is the energy density of sources, we have
\begin{align}\label{app}
U(t,{\bf x})-U_{\rm vac} = \, \, :U(t,{\bf x}):
\end{align}
where $: \cdot  :$ denotes normal-ordering of the creation and annihilation operators within $U_{\rm em}$.

By Poynting's theorem the local energy-flux can be written in terms of the Hermitian Poynting vector
\begin{align}\label{poynt}
{\bf S}(t,{\bf x})={1\over 2}\left[{\bf E}(t,{\bf x})\times {\bf B}(t,{\bf x}) +{\rm H.c.}\right].
\end{align}
Analogously to the vacuum energy-density we define the vacuum Poynting vector by
\begin{align}
S_{{\rm vac},i}&(t,{\bf x}) \nonumber \\ &= \langle S_{0,i}(t,{\bf x})\rangle_0  =\epsilon_{ijk} [E^{(+)}_{0,j}(t,{\bf x}),B^{(-)}_{0,k}(t,{\bf x})] \nonumber \\ &= \epsilon_{ijk} [E^{(+)}_j(t,{\bf x}),B^{(-)}_k(t,{\bf x})]
\end{align}
where $\epsilon_{ijk}$ denotes the Levi-Civita symbol. Since $U_{\rm vac}$ and ${\bf S}_{\rm vac}$ are independent of time and space Poynting's theorem can be written in terms of $:U(t,{\bf x}):\, =U(t,{\bf x})-U_{\rm vac}$ and $:{\bf S}(t,{\bf x}): \,={\bf S}(t,{\bf x})-{\bf S}_{\rm vac}$ as
\begin{align}\label{pt}
\partial_t \langle :U(t,{\bf x}):\rangle = -\nabla \cdot \langle :{\bf S}(t,{\bf x}):\rangle.
\end{align}
The practice of normally-ordering operator products in this way constitutes a basic application of Wick's theorem.

For a dipole at ${\bf 0}$ the radiated power can be obtained by integrating the right-hand-side of Eq.~(\ref{pt}) over a sphere centered at ${\bf 0}$ with radius $x$. Using the divergence theorem and subsequently taking the limit $x\to \infty$ the associated intensity is then seen to be
\begin{align}\label{Irad4}
&I_{\rm rad}(t,{\bf x})={\hat {\bf x}}\cdot \langle {\bf S}_{\rm rad}(t,{\bf x})-{\bf S}_{\rm vac}\rangle, \nonumber \\ & {\bf S}_{\rm rad}={1\over 2}\left[  {\bf E}_{\rm rad} \times {\bf B}_{\rm rad} + {\rm H.c.}\right]
\end{align}
where ${\bf E}_{\rm rad}(t,{\bf x}) = {\bf E}_0(t,{\bf x})+{\bf E}_{{\rm rad},s}(t,{\bf x})$ and ${\bf B}_{\rm rad}(t,{\bf x}) = {\bf B}_0(t,{\bf x})+{\bf B}_{{\rm rad},s}(t,{\bf x})$. Assuming the initial state $\ket{e,0}$ and making use of the identities
\begin{align}
&{\hat {\bf x}}\cdot {\bf E}_{{\rm rad},s}(t,{\bf x}) =0= {\hat {\bf x}}\cdot{\bf B}_{{\rm rad},s}(t,{\bf x}),\nonumber\\
&{\hat {\bf x}}\times {\bf E}_{{\rm rad},s}(t,{\bf x})={\bf B}_{{\rm rad},s}(t,{\bf x}),\nonumber \\ &
{\hat {\bf x}}\times {\bf B}_{{\rm rad},s}(t,{\bf x})=-{\bf E}_{{\rm rad},s}(t,{\bf x}),\nonumber \\ &
{\bm {\mathcal B}}^2_{\rm rad} = {\bm {\mathcal E}}^2_{\rm rad}
\end{align}
it is straightforward to show that the radiation intensity can also be written
\begin{align}\label{Irad3}
I_{\rm rad}(t,{\bf x}) =\langle {\bf E}_{\rm rad}(t,{\bf x})^2-{\bf E}_0(t,{\bf x})^2 \rangle_{e;0} = 2I_G(t,{\bf x})
\end{align}
where the second equality follows within the rotating-wave approximation. This explains why $2I_G$ rather than $I_G$ yields the correct radiated power. The anti-normally ordered combination ${\bf E}^{(+)}_{\rm rad}(t,{\bf x})\cdot {\bf E}^{(-)}_{\rm rad}(t,{\bf x})$, which is such that $\langle:{\bf E}^{(+)}_{\rm rad}(t,{\bf x})\cdot {\bf E}_{\rm rad}^{(-)}(t,{\bf x}):\rangle =\langle {\bf E}^{(-)}_{\rm rad}(t,{\bf x})\cdot {\bf E}^{(+)}_{\rm rad}(t,{\bf x}) \rangle$, is slowly-varying (resonant), so it is not eliminated by the rotating-wave and Markov approximations.

\subsection{Vacuum source-field correlations}\label{vsc}

We now turn our attention to the contribution of vacuum source-correlations to the radiation intensity. In references \cite{power_quantum_1992,salam_molecular_2008,salam_molecular_2009,stokes_quantum_2016} Eq.~(\ref{Irad4}) is used to perturbatively calculate the radiated power. We provide a simpler derivation of the same result, and in section \ref{np} we provide a non-perturbative extension. The physical intensity is determined relative to the vacuum value, so assuming the initial state $\ket{e,0}$ the intensity can be written according to Eq.~(\ref{Irad3}) as
\begin{align}\label{Irad2}
I_{\rm rad}(t,{\bf x}) = &\langle {\bf E}_{{\rm rad},s}(t,{\bf x})^2\rangle_{e,0} + \langle {\bf E}_0(t,{\bf x})\cdot {\bf E}_{{\rm rad},s}(t,{\bf x}) \nonumber \\ &+ {\bf E}_{{\rm rad},s}(t,{\bf x})\cdot {\bf E}_0(t,{\bf x})\rangle_{e,0}.
\end{align}

First we calculate the anti-normally ordered combination $\langle {\bf E}^{(+)}_{{\rm rad},s}(t,{\bf x})\cdot {\bf E}^{(-)}_{{\rm rad},s}(t,{\bf x})\rangle_{e;0}$, which is easily found using Eq.~(\ref{Epms}) to be
\begin{align}\label{Ian}
\langle {\bf E}^{(+)}_{{\rm rad},s}&(t,{\bf x})\cdot {\bf E}^{(-)}_{{\rm rad},s}(t,{\bf x})\rangle_{e;0} =\nonumber \\ &\left({1\over 4\pi x}\right)^2 \sum_{m>e} \omega_{em}^4  [{\bf d}_{em}-{\hat {\bf x}}({\hat {\bf x}}\cdot {\bf d}_{em})]^2
\end{align}
to second order in the transition dipole moments. By substituting Eqs.~(\ref{IG2}) and (\ref{Ian}) into Eq.~(\ref{Poynt}) we obtain, within the rotating-wave approximation, the contribution of ${\bf E}_{{\rm rad},s}^2$ in Eq.~(\ref{Irad2}) to the radiated power to second order as
\begin{align}\label{Ps}
P_s = \int d\Omega  \, x^2  \langle {\bf E}_{{\rm rad},s}(t,{\bf x})^2 \rangle = {1\over 2} \sum_m \omega_{em}\Gamma_{em}
\end{align}
where all dipole levels $m$ are included within the summation.

It remains to calculate the contribution of the vacuum-source correlations on the first line in Eq.~(\ref{Irad2}). This can be done a number of ways, with varying complexity. The authors \cite{power_quantum_1992,salam_molecular_2008,salam_molecular_2009} use the full source-field soultions of the electric and magnetic fields and separate out time-independent and oscillatory contributions to the power. The time-independent contributions give our final result Eq.~(\ref{Pvac}) below, while the time-dependent contributions vanish on average, and can be interpreted as virtual transients. Reference \cite{stokes_quantum_2016} gives a simplified derivation of Eq.~(\ref{Pvac}), which involves using only the radiation source-fields rather than the full source-fields. Here we give a yet simpler derivation which makes use of the Markov and rotating-wave approximations. The latter eliminate virtual photon contributions, and therefore yield Eq.~(\ref{Pvac}) directly, without requiring a time-average.

The vacuum-source correlation terms within (\ref{Irad2}), are found to second order to be (appendix \ref{ap2})
\begin{align}
\langle {\bf E}_0^{(+)}&(t,{\bf x})\cdot {\bf E}^{(-)}_{{\rm rad},s}(t,{\bf x})\rangle_{0;e} + {\rm c.c.}=\nonumber \\ & {1\over (4\pi x)^2} \sum_m {\rm sgn}(\omega_{em})\omega^4_{em}[{\bf d}_{em}-{\hat {\bf x}}({\hat {\bf x}}\cdot {\bf d}_{em})]^2
\end{align}
in agreement with references \cite{power_quantum_1992,salam_molecular_2008,salam_molecular_2009,stokes_quantum_2016}. The corresponding contribution to the radiated power is
\begin{align}\label{Pvac}
P_{\rm vac-s}&= \int d\Omega  \, x^2  \langle  {\bf E}_0(t,{\bf x})\cdot {\bf E}_{{\rm rad},s}(t,{\bf x}) \rangle+{\rm c.c.} \nonumber \\ &={1\over 2}\sum_m {\rm sgn}(\omega_{em}) \omega_{em}\Gamma_{em}.
\end{align} 
By adding the two contributions Eqs.~(\ref{Ps}) and (\ref{Pvac}) one obtains the correct result Eq.~(\ref{P}). Vacuum-source correlations are seen to be responsible for supplying half of the total radiated power, as well as for eliminating, in the final result, the contributions of energy non-conserving virtual transitions for which $m>e$.

\subsection{Non-perturbative extension}\label{np}

The results of section \ref{vsc} can be extended beyond the use of second order perturbation theory. For this purpose we restrict our attention to a two-level emitter. Judicious use of the Markov and rotating-wave approximations within the Heisenberg equations, leads to a tractable system of coupled Heisenberg-Langevin equations for the dipole operators, which can then be formally integrated.  

We begin by using the solution (\ref{a}), and the rotating-wave and Markov approximations to calculate the dipole's own reaction field as
\begin{align}\label{ast2}
{\bf D}^{(+)}_{{\rm T},s}(t,{\bf 0}) &= i\int d^3 k \sqrt{\omega\over 2(2\pi)^3}\sum_\lambda {\bf e}_\lambda({\bf k})  a_{\lambda,s}(t,{\bf k}) \nonumber \\ &= i{\Gamma \over 2}{ {\hat {\bf d}} \over d} \sigma^-(t).
\end{align}
Substituting Eq.~(\ref{ast}) into the Heisenberg equation for each of $\sigma^+$ and $\sigma^z$, making the rotating-wave approximation, and formally integrating the result yields the solutions
\begin{align}
&\sigma^+(t)= e^{(i\omega_0-\Gamma/2)t}\sigma^+ +\int_0^t dt' e^{(i \omega_0-\Gamma/2)(t-t')}\sigma^+_{\rm vac}(t')\label{sol1} \\ 
&\sigma^z(t) +1 = e^{-\Gamma t}[\sigma^z +1]+ \int_0^t dt' e^{-\Gamma (t-t')}\sigma^z_{\rm vac}(t').\label{sol2}
\end{align}
where
\begin{align}\label{sigzvac}
&\sigma^+_{\rm vac}(t) = i[{\bf d}\cdot {\bf D}^{(-)}_{{\rm T},0}(t,{\bf 0})]\sigma^z(t) \\ 
&\sigma^z_{\rm vac}(t) = -2i[{\bf d}\cdot {\bf D}^{(-)}_{{\rm T},0}(t,{\bf 0})]\sigma^-(t) + {\rm H.c.}.
\end{align}
In addition to the solutions to the Bloch equations (without driving), the solutions (\ref{sol1}) and (\ref{sol2}) also contain contributions from the quantum vacuum field.

Using Eqs.~(\ref{sol1}), (\ref{sol2}) and (\ref{Ema}) we find the Glauber intensity function in the state $\ket{0,e}$ to be
\begin{align}\label{IG3}
I_G(t) =  \left({\omega_0^2 \over 4\pi x}[{\bf d}-{\hat {\bf x}}({\hat {\bf x}}\cdot {\bf d})]\right)^2  e^{-\Gamma t_r}
\end{align}
The corresponding power is
\begin{align}\label{pg}
P_G(t) = {1\over 2} \omega_0 \Gamma e^{-\Gamma t_r}
\end{align}
whose short-time limit coincides with the power in Eq.~(\ref{Ppm}) restricted to two levels.

Next, following the same method as in section \ref{vsc} we find the anti-normally ordered combination $\langle {\bf E}^{(+)}_{{\rm rad},s}(t,{\bf x})\cdot {\bf E}^{(-)}_{{\rm rad},s}(t,{\bf x})\rangle_{e;0}$ to be
\begin{align}\label{Ian2}
\langle {\bf E}^{(+)}_{{\rm rad},s}&(t,{\bf x})\cdot {\bf E}^{(-)}_{{\rm rad},s}(t,{\bf x})\rangle_{e;0}=\nonumber \\ & \left({\omega_0^2 \over 4\pi x}[{\bf d}-{\hat {\bf x}}({\hat {\bf x}}\cdot {\bf d})]\right)^2  (1-e^{-\Gamma t_r}),
\end{align}
which when combined with Eq.~(\ref{IG3}) gives the total contribution of the source-field to the radiated power as
\begin{align}\label{ps}
P_s(t) = {1\over 2}\omega_0 \Gamma.
\end{align}
Finally, the vacuum-source correlations are found using Eqs.~(\ref{sol1}), (\ref{sol2}), (\ref{Ema}) and (\ref{Dfree}) to be (appendix \ref{ap2})
\begin{align}
\langle {\bf E}^{(+)}_0&(t,{\bf x})\cdot {\bf E}^{(-)}_{{\rm rad},s}(t,{\bf x}) \rangle_{0,e}+{\rm c.c.} \nonumber \\ &= \left({\omega_0^2 \over 4\pi x }[{\bf d} - {\hat {\bf x}}({\hat {\bf x}}\cdot {\bf d})]\right)^2 \left(2e^{-\Gamma t_r}-1\right).
\end{align}
The corresponding contribution to the power is
\begin{align}\label{pvs}
P_{\rm vac-s}(t) =\omega_0 \Gamma \left(e^{-\Gamma t_r} - {1\over 2}\right).
\end{align}
By adding Eqs.~(\ref{ps}) and (\ref{pvs}) we obtain the total radiated power
\begin{align}\label{p}
P(t) = \omega_0 \Gamma e^{-\Gamma t_r} =2P_G(t)
\end{align}
whose short time limit coincides with the power in Eq.~(\ref{P}) for a two-level system.

These results mirror the previous perturbative results in that neither Eq.~(\ref{pg}) nor Eq.~(\ref{ps}) yield the correct radiated power. Furthermore, since the pure source power in Eq.~(\ref{ps}) does not decay, it is unphysical for long times when considered on its own. As with the previous perturbative results only the combination of the individually non-physical contributions (\ref{ps}) and (\ref{pvs}), yields the correct physical power.

\section{Vacuum-source correlations and advanced waves}\label{vscaw}

In section \ref{vscri} we showed that a specific quadratic field functional, namely, the intensity functional is the difference between the quantum version of the classical definition and the quantum vacuum value. Accounting for vacuum-source correlations was essential in obtaining the correct physical results.

The most general quadratic field functionals are correlation functions defined in terms of the components of the electromagnetic field tensor at different points in spacetime. In a designated (laboratory) inertial frame all such correlations have the form $\langle X_i(t,{\bf x})Y_j(t',{\bf x}')\rangle$ where ${\bf X},~{\bf Y} = {\bf E},~{\bf B}$. In the rotating-wave approximation the terms $\langle X_i^{(+)}(t,{\bf x})Y_j^{(+)}(t',{\bf x}')\rangle$ and $\langle X_i^{(-)}(t,{\bf x})Y_j^{(-)}(t',{\bf x}')\rangle$ are taken as rapidly oscillating and are negligible. The corresponding Glauber-type correlation function $\langle X_i^{(-)}(t,{\bf x})Y_j^{(+)}(t',{\bf x}')\rangle$ is therefore related to $\langle X_i(t,{\bf x})Y_j(t',{\bf x}')\rangle$ by
\begin{align}\label{glau}
&G_{X_i Y_j}(t,{\bf x}|t',{\bf x}')=2\langle X_i^{(-)}(t,{\bf x})Y_j^{(+)}(t',{\bf x}')\rangle \nonumber \\ &\approx \langle X_i(t,{\bf x})Y_j(t',{\bf x}') - [X^{(+)}_i(t,{\bf x}),Y^{(-)}_j(t',{\bf x}')] \rangle.
\end{align}
In the vacuum state the commutator $[X^{(+)}_i(t,{\bf x}),Y^{(-)}_j(t',{\bf x}')]$, which is subtracted from $\langle X_i(t,{\bf x})Y_j^(t',{\bf x}')\rangle$  in Eq.~(\ref{glau}) includes the pure vacuum value $\langle X_{0,i}(t,{\bf x})Y_{0,j}(t',{\bf x}')\rangle_0$ as well as source-dependent components. If instead we subtract only the vacuum value from $\langle X_i(t,{\bf x})Y_j(t',{\bf x}')\rangle$ we obtain the alternative correlation function
\begin{align}\label{tt}
&C_{X_iY_j}(t,{\bf x}|t',{\bf x}') \nonumber \\ &= \langle X_i(t,{\bf x})Y_j(t',{\bf x}') - [X^{(+)}_{0,i}(t,{\bf x}),Y^{(-)}_{0,j}(t',{\bf x}')] \rangle.
\end{align}
For $t\neq t'$ the commutators between the total fields ${\bf E}^{(\pm)}$ and ${\bf B}^{(\pm)}$ are not equal to their counterparts defined in terms of the vacuum fields ${\bf E}^{(\pm)}_0$ and ${\bf B}^{(\pm)}_0$. As a result even within the rotating-wave approximation $G_{X_iY_j}$ and $C_{X_iY_j}$ do not generally coincide. Unlike $G_{X_iY_j}$ the vacuum-fields contribute to $C_{X_iY_j}$ within the vacuum state. 

We define the difference operator
\begin{align}
\Delta_{X_i Y_j}(t,{\bf x}|t',{\bf x}') =& [X^{(+)}_i(t,{\bf x}),Y^{(-)}_j(t',{\bf x}')] \nonumber \\ &- [X^{(+)}_{0,i}(t,{\bf x}),Y^{(-)}_{0,j}(t',{\bf x}')]
\end{align}
in terms of which, in the rotating-wave approximation, we have
\begin{align}\label{CXY2}
C_{X_iY_j} = G_{X_iY_j}+\langle \Delta_{X_i Y_j}\rangle.
\end{align}

A perturbative calculation of certain correlation functions of the form given in Eq.~(\ref{tt}) in the particular case that $t=t'$ has been carried out without using the rotating-wave or Markov approximations within the non-relativistic QED framework by Power and Thirunamachandran \cite{power_quantum_1992,power_quantum_1993} (see also \cite{salam_molecular_2008,salam_molecular_2009,stokes_quantum_2016}). Their results are consistent with the results presented in section \ref{vsc}, and differ only by the inclusion of oscillatory transient contributions. The results in references \cite{power_quantum_1992,power_quantum_1993} are also consistent with the short-time limiting behaviour of the non-perturbative results given in section \ref{np}.

Here we consider the the more general case in which $t$ and $t'$ are arbitrary, which is fundamentally different, because $\Delta_{X_i Y_j}(t,{\bf x}|t',{\bf x}')$ is generally non-vanishing. While we utilise the rotating-wave and Markov approximations our approach is not perturbative and is similar to that used in section \ref{np}. More precisely, we make use of the Heisenberg-Lengavin equations for a two-level dipole. Details of the calculations leading to the results below are given in appendix \ref{ap3}.

Using Eq.~(\ref{Ema}) we write ${\bf X}^{(-)}_s(t,{\bf x}) = {\bm {\mathcal X}}({\bf x})\sigma^+(t-x)$ where ${\bf X}={\bf E},~{\bf B}$ and correspondingly ${\bm {\mathcal X}}={\bm {\mathcal E}},~{\bm {\mathcal B}}$. The $G_{X_iY_j}$ are easily found to be
\begin{align}\label{Gfin0}
&G_{X_iY_j}(t,{\bf x}|t',{\bf x}') =\nonumber \\ & 2{\mathcal X}_i({\bf x}){\mathcal Y}^*_j({\bf x}')\theta(t_r)\theta(t_r')\langle \sigma^+(t_r)\sigma^-(t_r')\rangle.
\end{align}
where $t_r = t- x$ and $t_r'=t'- x'$ and where $\theta$ denotes the Heaviside step-function. Assuming the initial state $\ket{e,0}$ we obtain
\begin{align}\label{Gfin}
&G_{X_iY_j}(t,{\bf x}|t',{\bf x}') =\nonumber \\ & 2{\mathcal X}_i({\bf x}){\mathcal Y}^*_j({\bf x}')\theta(t_r)\theta(t_r')e^{-[\Gamma/2](t_r+t_r')}e^{i\omega_0(t_r-t_r')}.
\end{align}
Next we consider $\Delta_{X_i Y_j}$, which can be partitioned into three terms as
\begin{align}
&\Delta_{X_i Y_j}(t,{\bf x}|t',{\bf x}') = [X^{(+)}_{s,i}(t,{\bf x}),Y^{(-)}_{s,j}(t',{\bf x}')] \nonumber \\ &+ [X^{(+)}_{0,i}(t,{\bf x}),Y^{(-)}_{s,j}(t',{\bf x}')]+[X^{(+)}_{s,i}(t,{\bf x}),Y^{(-)}_{0,j}(t',{\bf x}')].
\end{align}
The pure source component is easily found using Eq.~(\ref{Ema}) to be
\begin{align}\label{DXYs}
[&X^{(+)}_{s,i}(t,{\bf x}),Y^{(-)}_{s,j}(t',{\bf x}')] \nonumber \\ &= {\cal X}^*_i({\bf x}){\cal Y}_j({\bf x}')\theta(t_r)\theta(t_r')[\sigma^-(t_r),\sigma^+(t_r')].
\end{align}

The remaining components involve both the vacuum and source-fields, and they each consist of both retarded and advanced parts. After fairly lengthy calculations we obtain the compact expression (appendix \ref{ap3})
\begin{align}\label{retadv1} 
&[X^{(+)}_{0,i}(t,{\bf x}),Y^{(-)}_{s,j}(t',{\bf x}')] = \nonumber \\ &-{\cal X}^*_i({\bf x}){\cal Y}_j({\bf x}')\theta(t_r'-t_r)\theta(t_r)\theta(t_r')[\sigma^-(t_r),\sigma^+(t_r')]\nonumber \\ &\pm{\cal X}_i({\bf x}){\cal Y}_j({\bf x}')\theta(t_r'-t_a)\theta(t'_r)[\sigma^-(t_a),\sigma^+(t_r')],
\end{align}
where $t_a = t+ x$ and $t_a'=t'+ x'$, and the notation $\pm$ means that $+$ holds for the case ${\bf X}={\bf E}$ while $-$ holds when ${\bf X}={\bf B}$. Similarly we find that
\begin{align}\label{retadv2}
&[X^{(+)}_{s,i}(t,{\bf x}),Y^{(-)}_{0,j}(t',{\bf x}')] = \nonumber \\ &-{\cal X}^*_i({\bf x}){\cal Y}_j({\bf x}')\theta(t_r-t_r')\theta(t_r)\theta(t_r')[\sigma^-(t_r),\sigma^+(t_r')] \nonumber \\ &\pm {\cal X}^*_i({\bf x}){\cal Y}^*_j({\bf x}')\theta(t_r-t_a')\theta(t_r)[\sigma^-(t_r),\sigma^+(t_a')]
\end{align}
where $+$ holds for the case ${\bf Y}={\bf E}$ while $-$ holds when ${\bf Y}={\bf B}$. The sum of the purely retarded parts from Eqs.~(\ref{retadv1}) and (\ref{retadv2}) exactly cancels the pure source component given in Eq.~(\ref{DXYs}). We therefore obtain the final result
\begin{align}\label{DXYfin}
&\Delta_{X_i Y_j}(t,{\bf x}|t',{\bf x}') = \nonumber \\ &(-1)^\alpha  {\cal X}_i({\bf x}){\cal Y}_j({\bf x}')\theta(t_r'-t_a)\theta(t'_r)[\sigma^-(t_a),\sigma^+(t_r')]\nonumber \\ + &(-1)^\beta  {\cal X}^*_i({\bf x}){\cal Y}^*_j({\bf x}')\theta(t_r-t_a')\theta(t_r)[\sigma^-(t_r),\sigma^+(t_a')].
\end{align}
where $\alpha =0$ for ${\bf X}={\bf E}$ and $\alpha=1$ for ${\bf X}={\bf B}$ while $\beta =0$ for ${\bf Y}={\bf E}$ and $\beta=1$ for ${\bf Y}={\bf B}$. The Heaviside functions in this expression imply that $\Delta_{X_iY_j}(t,{\bf x}|t,{\bf x}')=0$ as required. However, for $t\neq t'$, $\langle \Delta_{X_iY_j}(t,{\bf x}|t',{\bf x}')\rangle$ generally consists of non-vanishing two-time correlations, which are due to the advanced components of the commutators of the free and source-fields in Eqs.~(\ref{retadv1}) and (\ref{retadv2}). 
%
%
\begin{figure}[t]
\begin{minipage}{\columnwidth}
\begin{center}
\hspace*{-0.6cm}\includegraphics[scale=1.05]{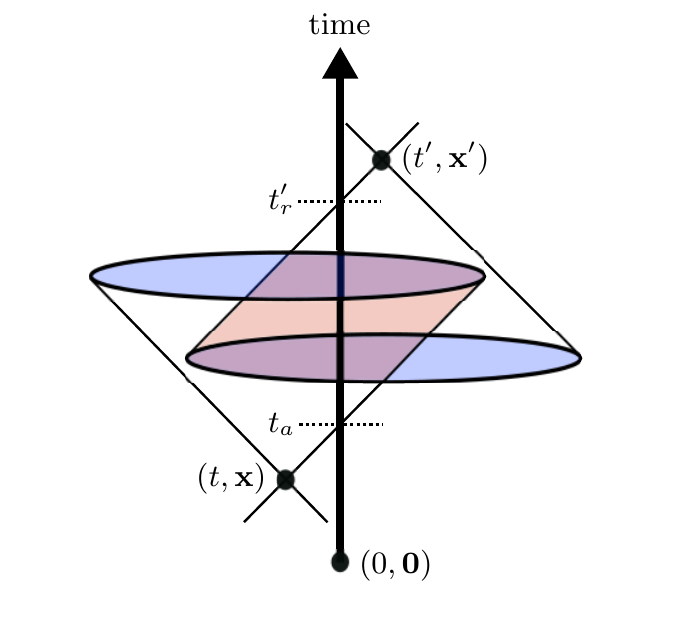} 
\vspace*{-0.7cm}
\caption{The world line of the dipole, represented by the central vertical arrow, passes through the intersection of the backward lightcone of the event $(t',{\bf x}')$ and the forward lightcone of $(t,{\bf x})$. The commutator $[X^{(+)}_{0,i}(t,{\bf x}),Y^{(-)}_{s,j}(t',{\bf x}')]$ contains a component that depends on the source operator $\sigma^-(t_a)$ where $t_a$ is the advanced time associated with the event $(t,{\bf x})$ at which the vacuum field $X^{(+)}_{0,i}(t,{\bf x})$ is evaluated.}\label{f}
\end{center}
\end{minipage}
\end{figure}
%
%
Assuming the initial state $\ket{e,0}$, with some work we obtain (appendix \ref{ap3})
\begin{align}\label{DXYexp}
\langle & \Delta_{X_i Y_j}(t,{\bf x}| t',{\bf x}')\rangle_{e,0} =\nonumber \\ & (-1)^\alpha{\cal X}_i({\bf x}){\cal Y}_j({\bf x}')\theta(t_r'-t_a)\theta(t'_r)\theta(t_a)\nonumber \\& ~~~\times e^{(i\omega_0-\Gamma/2)t_r'}e^{(-i\omega_0-\Gamma/2)t_a}(e^{\Gamma t_a}-2) \nonumber \\ &+ (-1)^\beta{\cal X}^*_i({\bf x}){\cal Y}^*_j({\bf x}')\theta(t_r-t_a')\theta(t_r)\theta(t_a')\nonumber \\ &~~~\times e^{(-i\omega_0-\Gamma/2)t_r}e^{(i\omega_0-\Gamma/2)t_a'}(e^{\Gamma t_a'}-2).
\end{align}

Since $\Delta_{X_i,Y_j}(t,{\bf x}|t',{\bf x}')$ is non-local in both space and time the presence of advanced contributions is not acausal. Supposing for definiteness that $t'>t$ then all fields at events within the backward lightcone of $(t',{\bf x}')$ may contribute to $\Delta_{X_i,Y_j}(t,{\bf x}|t',{\bf x}')$. As figure \ref{f} illustrates, provided a portion of the dipole's world line belongs to the intersection of the backward lightcone of $(t',{\bf x}')$ and the forward lightcone of $(t,{\bf x})$ we have $t_r'\geq t_a$. In this case $\Delta_{X_i,Y_j}(t,{\bf x}|t',{\bf x}')$ receives a non-zero advanced contribution coming from the first line in Eq.~(\ref{DXYfin}).

\section{Photo-detection theory}\label{pd}

The theory of photo-detection is fundamental within quantum optics and was initiated some time ago \cite{mandel_fluctuations_1958,mandel_theory_1964,glauber_quantum_1963}. In this section we briefly verify that when source and detector are taken as identical two-level dipoles, advanced wave correlations do not contribute to photo-detection amplitudes within the rotating-wave and Markov approximations. 

The source is located at the origin ${\bf 0}$ and the detector is located at a position ${\bf x}$ with $x\gg \omega_0^{-1}$. The interaction Hamiltonian for the model is given by
\begin{align}
V=-{\hat {\bf d}}_s \cdot {\bf D}_{\rm T}({\bf 0})-{\hat {\bf d}}_d \cdot {\bf D}_{\rm T}({\bf x})
\end{align}
where the subscripts $s$ and $d$ stand for source and detector respectively. The state of the composite system in which the source is in the excited state, the detector is in the ground state, and the field contains no photons is denoted $\ket{\psi}$. The rate of excitation of the detector is found using the equation of motion for $\langle \sigma_d^z(t)\rangle_\psi$ to be
\begin{align}\label{sigzd}
{\dot p}_d(t) =& -d_i d_j \int_0^t dt'  e^{-i\omega_0(t-t')} \nonumber \\ &\times  \langle D_{{\rm T},i}(t,{\bf x})  D_{{\rm T},j}(t',{\bf x})\sigma_d^z(t')\rangle_\psi  +{\rm c.c.}\, .
\end{align}
Since actual photo-detection events depend on photo-ionisation of an electron within the continuum of final detector states, it is argued in \cite{milonni_photodetection_1995} that the rate of stimulated emission from an excited state should be assumed to be much smaller than the rate of absorption from the ground state. This assumption allows us to make the perturbative approximation $\sigma^z_d(t') \approx \sigma_d^z(0)$ in Eq.~(\ref{sigzd}), which for a detector in the ground state yields
\begin{align}\label{sigzd2}
{\dot p}_d(t)=& d_i d_j \int_0^t dt' \langle D_{{\rm T},i}(t,{\bf x})  D_{{\rm T},j}(t',{\bf x}) \rangle_\psi e^{-i\omega_0(t-t')} \nonumber \\ &+{\rm c.c.}\,.
\end{align}
The total transverse displacement field at the position of the detector includes the detectors own reaction field, and is given by
\begin{align}\label{D2}
{\bf D}_{\rm T}(t,{\bf x}) = {\bf E}_0(t,{\bf x}) + {\bf E}_s(t,{\bf x}) + {\bf D}_{{\rm T},d}(t,{\bf x}).
\end{align}

Using again the perturbative approximation $\sigma^\pm_d(t)\approx \sigma^\pm_d e^{\pm i\omega_0 t}$ for the detector operators, and the rotating-wave approximation, all terms in Eq.~(\ref{sigzd2}) that involve the detector's own reaction field are seen to vanish under the assumption that the detector is in the ground state. Moreover, while the pure vacuum correlation function $\langle E_{0,i}^{(+)}(t,{\bf x})  E_{0,j}^{(-)}(t',{\bf x}) \rangle_\psi$ can be shown to contribute to the detector level-shifts, it does not contribute within the rotating-wave and Markov approximations to the detectors excitation \cite{milonni_photodetection_1995}. We therefore obtain
\begin{align}\label{sigzd3}
{\dot p}_d(t)= d_i d_j \int_0^t dt'  C_{E_iE_j}(t,{\bf x}|t',{\bf x}) e^{-i\omega_0(t-t')} +{\rm c.c.}
\end{align}
The advanced contributions to this expression are found using Eq.~(\ref{DXYexp}) with ${\bf x}={\bf x}'$. Since $t'$ takes values from $0$ upto $t>0$ the first term on the right-hand-side of Eq.~(\ref{DXYexp}) does not contribute. It now suffices to note that the second term on the right-hand-side of Eq.~(\ref{DXYexp}) possesses an oscillatory dependence of the form $e^{-i\omega_0(t-t')}$, so that when substituted into Eq.~(\ref{sigzd3}) this term gives only rapidly oscillating contributions to ${\dot p}_d(t)$, which are negligible using the rotating-wave approximation. Within the rotating wave approximation we therefore have that
\begin{align}\label{sigzd4}
{\dot p}_d(t)= d_i d_j \int_0^t dt'  G_{E_iE_j}(t,{\bf x}|t',{\bf x}) e^{-i\omega_0(t-t')} +{\rm c.c.},
\end{align}
which is essentially the same result originally derived by Glauber \cite{glauber_quantum_1963}.

A few concluding remarks are in order regarding the final result Eq.~(\ref{sigzd4}). We note that the replacement of $C_{E_iE_j}$ by $G_{E_iE_j}$ in the last step is a result of the rotating-wave approximation. Therefore, advanced wave components of $C_{E_iE_j}$ may in principle contribute to the detector excitation, but such contributions are no larger than conventional counter-rotating contributions involving, for example, the detector's own reaction field.

The rotating-wave approximation is generally considered to be valid for optical frequencies and realistic detector response times \cite{milonni_photodetection_1995}. On the other hand, in the more realistic case of multi-level non-identical dipoles, the validity of the rotating-wave approximation is less clear. One could also consider detection on much shorter time scales. However, it is not entirely clear how regimes which require moving beyond the rotating-wave approximation should be modelled. In such regimes a retention of the conventional ontology consisting of atomic excitations and photons defined in terms of a bare energy $H_0$, which is not even approximately conserved, seems physically dubious. Little more can be said presently about the role of advanced waves within such regimes. We remark that the techniques found in \cite{drummond_unifying_1987,baxter_gauge_1990,stokes_extending_2012}, which allow one to eliminate counter-rotating terms from the dipole-field interaction Hamiltonian may offer a possible recourse. 

\section{Radiative forces}\label{forces}

In this section we show that certain measurement statistics involving the radiative force acting on an elementary quantum point charge $q$ due to the field of a quantum dipole, depend on the two-time correlation functions found in section \ref{vscaw}. The point charge has position and momentum operators ${\bf r}$ and ${\bf p}$. The Lorentz force ${\bf F}(t)=m{\ddot {\bf r}}(t)$ experienced by the charge due to an electromagnetic field is
\begin{align}\label{ext}
m{\ddot {\bf r}}(t) =& q{\bf E}(t,{\bf r}(t))\nonumber \\ &+{q\over 2}[{\bf {\dot r}}(t)\times {\bf B}(t,{\bf r}(t))-{\bf B}(t,{\bf r}(t))\times {\bf {\dot r}}(t)].
\end{align}
The momentum ${\bm \pi} = m{\dot {\bf r}}$ is found by integrating Eq.~(\ref{ext}). As an example of a statistical quantity that depends on two-time correlations in the external fields we consider the rate of change of the dispersion of the momentum, i.e., the momentum diffusion, which is
\begin{align}\label{disp}
&{d \over dt} \Delta{\bm \pi}(t) = (\langle {\bm \pi}(0) \cdot {\bf F}(t)\rangle +{\rm c.c.})-2\langle {\bm \pi}(0)\rangle \cdot \langle{\bf F}(t)\rangle \nonumber \\ &~+\int_0^t dt' \big(\left[\langle {\bf F}(t)\cdot {\bf F}(t')\rangle +{\rm c.c.}\right]  - 2\langle {\bf F}(t) \rangle \cdot \langle {\bf F}(t')\rangle\big).
\end{align}
Substitution of the Lorentz force given by Eq.~(\ref{ext}) into Eq.~(\ref{disp}), leads to a dependence of the momentum diffusion on integrated two-time correlation functions of the electromagnetic field. Unlike in photo-detection theory, these correlation functions are not accompanied by oscillatory factors due to the detector itself. As a result the advanced components of the relevant correlation functions are not generally negligible.

We now derive an explicit expression for the momentum diffusion based on a series of simplifying assumptions and approximations. We begin by noting that a free charge $q$ can be described using the Coulomb-gauge Hamiltonian
\begin{align}\label{hq}
H' = {1\over 2m}[{\bf p}-q{\bf A}_{\rm T}({\bf r})]^2 + V_{\rm self}({\bf r}) + H_f
\end{align}
where ${\bf A}_{\rm T}$ denotes the transverse vector potential, $V_{\rm self}$ denotes the Coulomb self-energy of the charge, and the Coulomb gauge free field Hamiltonian is defined in terms of the transverse electric field ${\bf E}_{\rm T} = -{\dot {\bf A}}_{\rm T}$ as
\begin{align}
H_f ={1\over 2}\int d^3 x \, [{\bf E}_{\rm T}({\bf x})^2 +{\bf B}({\bf x})^2].
\end{align}
The total electric field is found by adding to ${\bf E}_{\rm T}$ the longitudinal field ${\bf E}_{\rm L}=-\nabla \phi_{\rm coul}$ where $\phi_{\rm coul}$ is the Coulomb potential of the point charge. The vacuum and source components of the transverse electric field ${\bf E}_{{\rm T},0}$ and ${\bf E}_{{\rm T},s}$, which are defined by ${\bf E}_{\rm T}={\bf E}_{{\rm T},0}+{\bf E}_{{\rm T},s}$ are such that changes in ${\bf E}_{{\rm T},0}$ are generated by the Hamiltonian $H_f$ alone, i.e., ${\dot {\bf E}}_{{\rm T},0}(t,{\bf x}) =-i[{\bf E}_{{\rm T},0}(t,{\bf x}),H_f(0)]$. The source component ${\bf E}_{{\rm T},s}(t,{\bf x})$ includes two separate acausal components (appendix \ref{ap4}). The first is equal to $-{\bf E}_{\rm L}(t,{\bf x})$, which is independent of the definition of $H_f$ and which is cancelled out within the source-field ${\bf E}_s(t,{\bf x})$. The second instantaneous component depends on the source at the initial time $t=0$, and arises because in the Coulomb gauge $H_f$ is defined in terms of ${\bf E}_{\rm T}$ rather than ${\bf E}$ itself. Due to this component the electric source-field ${\bf E}_s$ is not causal within the Coulomb gauge. A complete derivation of the Coulomb gauge electric source-field is given in appendix \ref{ap4}.

The occurrence of a static precursor within the electric source-field also arises in the Coulomb gauge treatment of a dipole system \cite{power_time_1999}. On the other hand, in the multipolar gauge the free field Hamiltonian $H_f$ is defined in terms of the transverse displacement field ${\bf D}_{\rm T}$, which possesses a fully retarded source component and is equal to the total electric field away from the dipole \cite{cohen-tannoudji_photons_1997,power_time_1999}. Using superscripts $C$ and $m$ to refer to the Coulomb and multipolar gauges respectively, we have that although ${\bf E}^C_0\neq {\bf E}^m_0$ and ${\bf E}_s^C\neq {\bf E}_s^m$ it can be shown that ${\bf E}_0^C-{\bf E}^m_0 = {\bf E}_s^m-{\bf E}_s^C$, so that the total field ${\bf E}$ is indeed gauge-invariant \cite{power_time_1999}. Only the multipolar gauge however, correctly identifies the physical vacuum and source components ${\bf E}_0$ and ${\bf E}_s$, such that ${\bf E}_s$ is fully causal and is the quantised version of the classical field of an oscillating dipole as given in Eq.~(\ref{Erad}).

To correctly identify the vacuum and source components of the electromagnetic fields when considering the free charge $q$, we seek a description analogous to the multipolar description of the dipole. To this end we consider a heavy charge with relatively slow dynamics characterised by the time scale $t_q \gg \Gamma^{-1}$. We assume a state of the charge such that initially the spread in position is small compared with an optical wavelength $\sqrt{\Delta{\bf r}} \ll \omega_0^{-1}$, and such that the spread in Doppler shifts is small compared to the dipole's decay rate; $\omega_0 \sqrt{\Delta {\bf p}}/m \ll \Gamma$. The initial state of the charge is therefore well-localised in position and momentum.

Since the coupling of the charge to the field is weak and $t\ll t_q$ we have that ${\bf A}_{\rm T}(t,{\bf r}(t))\approx {\bf A}_{\rm T}(t,{\bf r}(0)+{\bf p}(0)t/m)$. Furthermore since the atom is initially well-localised in position and momentum we have  ${\bf A}_{\rm T}(t,{\bf r}(0)+{\bf p}(0)t/m)\approx {\bf A}_{\rm T}(t,{\bf r}_0+{\bf p}_0t/m)$ where ${\bf r}_0=\langle {\bf r}(0)\rangle$ and ${\bf p}_0=\langle {\bf p}(0)\rangle$. Finally we restrict our attention to a charge for which ${\bf p}_0 ={\bf 0}$, which implies that ${\bf A}_{\rm T}(t,{\bf r}(t))\approx {\bf A}_{\rm T}(t,{\bf r}_0)$. In summary, for times $t\ll t_q$ the initial spatial localisation of the charge is maintained, which means that $\langle {\bf r}(t)\rangle -{\bf r}_0$ remains small compared with $\omega_0^{-1}$. For times $t \ll t_q$ the kinetic energy of the charge is therefore
\begin{align}
H_q(t) \approx {1\over 2m}[{\bf p}(t)-q{\bf A}_{\rm T}(t,{\bf r}_0)]^2.
\end{align}
In this Hamiltonian the charge's initial wave-packet centre ${\bf r}_0$ acts like the reference centre of a dipole, which allows us to construct a description of the charge analogous to the multipolar description of a dipole . Specifically, we are now able to transform out the vector potential dependence of $H_q(t)$ via a unitary transformation $e^{-iq {\bf r}(t)\cdot {\bf A}_{\rm T}(t,{\bf r}_0)}$. 

We transform the Coulomb gauge Hamiltonian for the whole system consisting of the charge, dipole and field using the unitary $e^{-iq {\bf r}(t)\cdot {\bf A}_{\rm T}(t,{\bf r}_0)}e^{-i{\bf d}(t)\cdot {\bf A}_{\rm T}(t,{\bf 0})}$. We then make the two-level and rotating-wave approximations for the dipole and in the usual way, we absorb all dipole and charge self-energy terms into the definitions of their respective bare energies. This yields the final Hamiltonian $H=H_0+V$ where
\begin{align}\label{dqf}
&H_0 \nonumber \\ &= \omega_0\sigma^+\sigma^- + {{\bf p}^2\over 2m} + \int d^3k \sum_\lambda \omega \left(a_\lambda^\dagger ({\bf k}) a_\lambda ({\bf k})+{1\over 2}\right), \nonumber \\ 
&V = \, {\bf d}\cdot  {\bf \Pi}^{(-)}_{\rm T}({\bf 0})\sigma^- + \sigma^+{\bf d}\cdot  {\bf \Pi}^{(+)}_{\rm T}({\bf 0}) + q{\bf r}\cdot {\bf \Pi}_{\rm T}({\bf r}_0)
\end{align}
with
\begin{align}
&{\bf \Pi}_{\rm T}^{(+)}(t,{\bf x}) \nonumber \\ &= -i\int d^3 k \sum_\lambda \sqrt{\omega \over 2(2\pi)^3} {\bf e}_\lambda({\bf k}) a_\lambda(t,{\bf k})e^{i{\bf k}\cdot {\bf x}}.
\end{align}
The field canonical momentum ${\bf \Pi}_{\rm T}$ coincides with minus the total electric field away from the dipole and the charge; ${\bf \Pi}_{\rm T}({\bf x})=-{\bf E}({\bf x}),~{\bf x}\neq {\bf 0},{\bf r}_0$. In the description that uses the Hamiltonian in Eq.~(\ref{dqf}) the electric source-field is a properly causal field, and moreover, the charge's canonical momentum ${\bf p}$ coincides with the mechanical momentum $m{\dot {\bf r}}$.

Since we can replace ${\bm \pi}(0)$ by ${\bf p}_0={\bf 0}$ in the first term in Eq.~(\ref{disp}) this term is seen to vanish. To obtain the momentum diffusion correct to second order in $q$ we require the force ${\dot {\bf p}}(t)$ correct to first order in $q$, which is found using the Hamiltonian (\ref{dqf}) to be
\begin{align}\label{eext}
{\dot {\bf p}}(t) = q{\bf E}_e(t,{\bf r}_0),\qquad  {\bf E}_e = {\bf E}_0 + {\bf E}_s,
\end{align}
where ${\bf E}_0$ and ${\bf E}_s$ are given in Eqs.~(\ref{Dfree}) and (\ref{Ema}) respectively. By only considering terms upto order $q$ we find that the charge does not influence the dipole source operators on which the external field in Eq.~(\ref{eext}) depends. We may therefore also use Eqs.~(\ref{sol1}) and (\ref{sol2}). Assuming the initial state $\ket{e,0}$ for the dipole and the radiation field, the momentum diffusion is to $O(q^2)$ found using Eq.~(\ref{eext}) to be
\begin{align}\label{disp3}
{d \over dt} \Delta{\bf p}(t) = q^2\int_0^t dt' \langle {\bf E}_e(t,{\bf r}_0) \cdot {\bf E}_e(t',{\bf r}_0)\rangle_{0,e} +{\rm c.c.}
\end{align}
The correlation function in this expression includes an infinite pure-vacuum contribution which is independent of the dipole. A similar contribution was found in the context of photo-detection theory in section \ref{pd} where it was found to be equivalent to the detectors own radiation-reaction. This can be understood as a fluctuation-dissipation relation \cite{milonni_photodetection_1995}. Neglecting the pure-vacuum contribution in Eq.~(\ref{disp3}) while retaining all components that depend on the source dipole we obtain
\begin{align}\label{disp4}
{d \over dt} \Delta{\bf p}(t) = q^2\int_0^t dt' C_{E_iE_i}(t,{\bf r}_0|t',{\bf r}_0) +{\rm c.c.}
\end{align}
where $C_{E_iE_i}$ is defined in Eq.~(\ref{tt}). Using Eq.~~(\ref{CXY2}) we write Eq.~(\ref{disp4}) as the sum of two contributions; $d\Delta{\bf p}/dt=d \Delta{\bf p}_s/dt +d\Delta{\bf p}_{{\rm vac}-s}/dt$. Assuming the initial state $\ket{e,0}$ and restricting our attention to the radiation zone $x\gg \omega_0^{-1}$, the first contribution is found using Eq.~(\ref{Gfin}) to be
\begin{align}\label{m1}
{d \over dt}& \Delta{\bm p}_s(t) = q^2 \int_0^t dt' \, G_{E_iE_i}(t,{\bf r}_0 |t',{\bf r}_0) +{\rm c.c.} \nonumber \\ =& {2\theta(t_r)\over {\cal N}}\nonumber \\ &\times \big[e^{-{\Gamma\over 2}t_r} \left(\Gamma \cos[\omega_0t_r] +2\omega_0 \sin[\omega_0t_r] \right) -\Gamma e^{-\Gamma t_r}\big]
\end{align}
where
\begin{align}
{1\over {\cal N}}=  {q^2{\bm {\mathcal E}}_{\rm rad}({\bf r}_0)^2\over (\Gamma/2)^2 + \omega_0^2}
\end{align}
with ${\bm {\mathcal E}}_{\rm rad}$ the radiative component of ${\bm {\mathcal E}}$ given in Eq.~(\ref{ebcoeffs}). What remains is the advanced-wave contribution arising due to vacuum source-field correlations. This contribution is found using Eq.~~(\ref{DXYexp}) to be
\begin{align}\label{m2}
{d \over dt}& \Delta{\bf p}_{{\rm vac}-s}(t)=q^2\int_0^t dt' \, \langle \Delta_{E_iE_i}(t,{\bf r}_0 |t',{\bf r}_0)\rangle +{\rm c.c.} \nonumber \\ =& \, {1\over {\cal N}}\theta(t_r-r_0) \bigg[\Gamma(2e^{-\Gamma t_r}+1) \nonumber \\ &- e^{-{\Gamma\over 2}t} \big(\Gamma(e^{\Gamma r_0}+2)\cos[\omega_0(t-2r_0)] \nonumber \\ &~~~~-2\omega_0(e^{\Gamma r_0}-2)\sin[\omega_0(t-2r_0)] \big)\bigg].
\end{align}
Comparing Eqs.~(\ref{m1}) and (\ref{m2}) we see that the contribution from $G_{E_iE_i}$ is non-zero for times $t\geq r_0$ while the contribution from $\langle \Delta_{E_iE_i}\rangle$ is only non-zero for times $t\geq 2r_0$. This allows the two contributions to be clearly distinguished.
%
%
\begin{figure}[t]
\begin{minipage}{\columnwidth}
\begin{center}
\hspace*{-2cm}\includegraphics[scale=0.82]{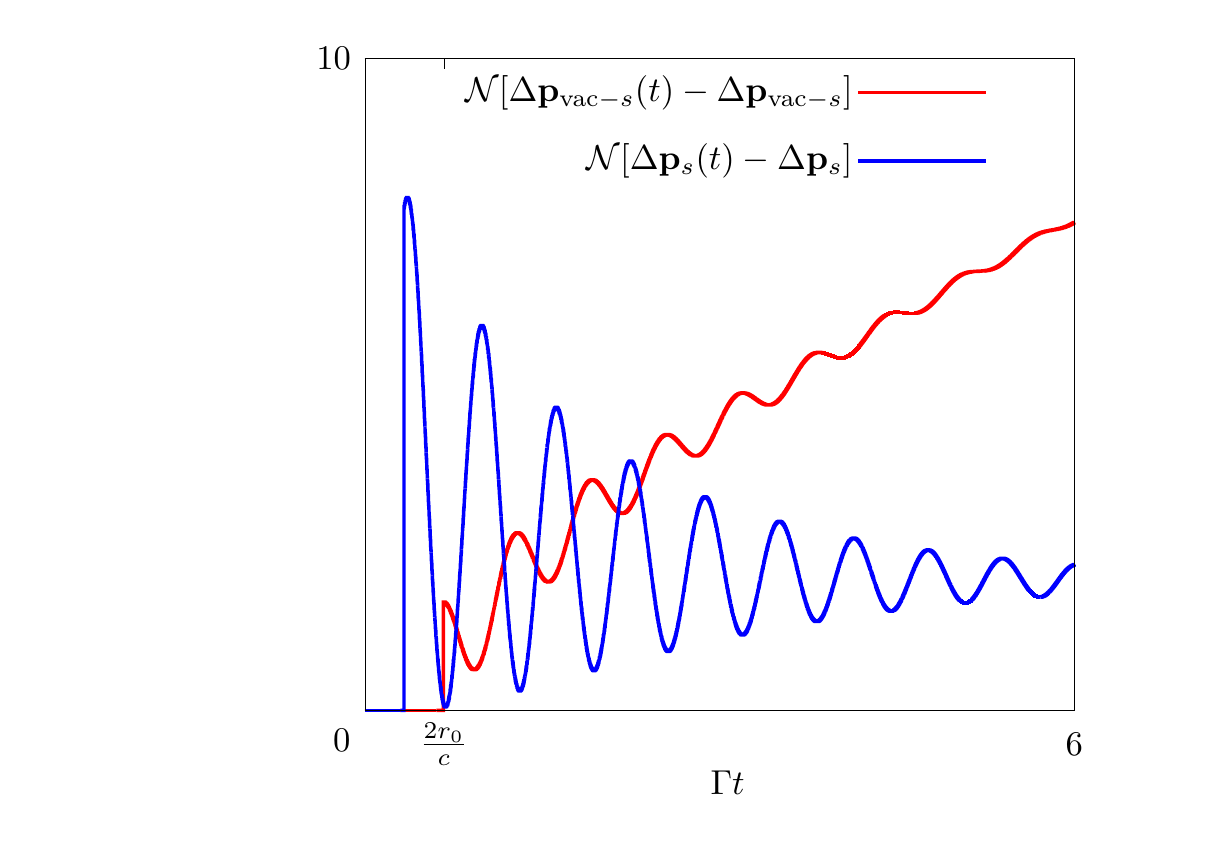} 
\vspace*{-1cm}
\caption{The contributions ${\cal N}[\Delta{\bf p}_s(t)-\Delta{\bf p}_s]$ and ${\cal N}[\Delta{\bf p}_{{\rm vac}-s}(t)-\Delta{\bf p}_{{\rm vac}-s}]$ are plotted. The decay rate $\Gamma=10^8$ is chosen in the optical regime. To clearly illustrate the oscillatory character the transition frequency has been chosen as $\omega_0=10\Gamma$. The time $r_0/c$ where $c$ denotes the speed of light in the vacuum has been chosen as $1/(3\Gamma)$.}\label{f1}
\end{center}
\end{minipage}
\end{figure}
\begin{figure}[t]
\begin{minipage}{\columnwidth}
\begin{center}
\hspace*{-2cm}\includegraphics[scale=0.85]{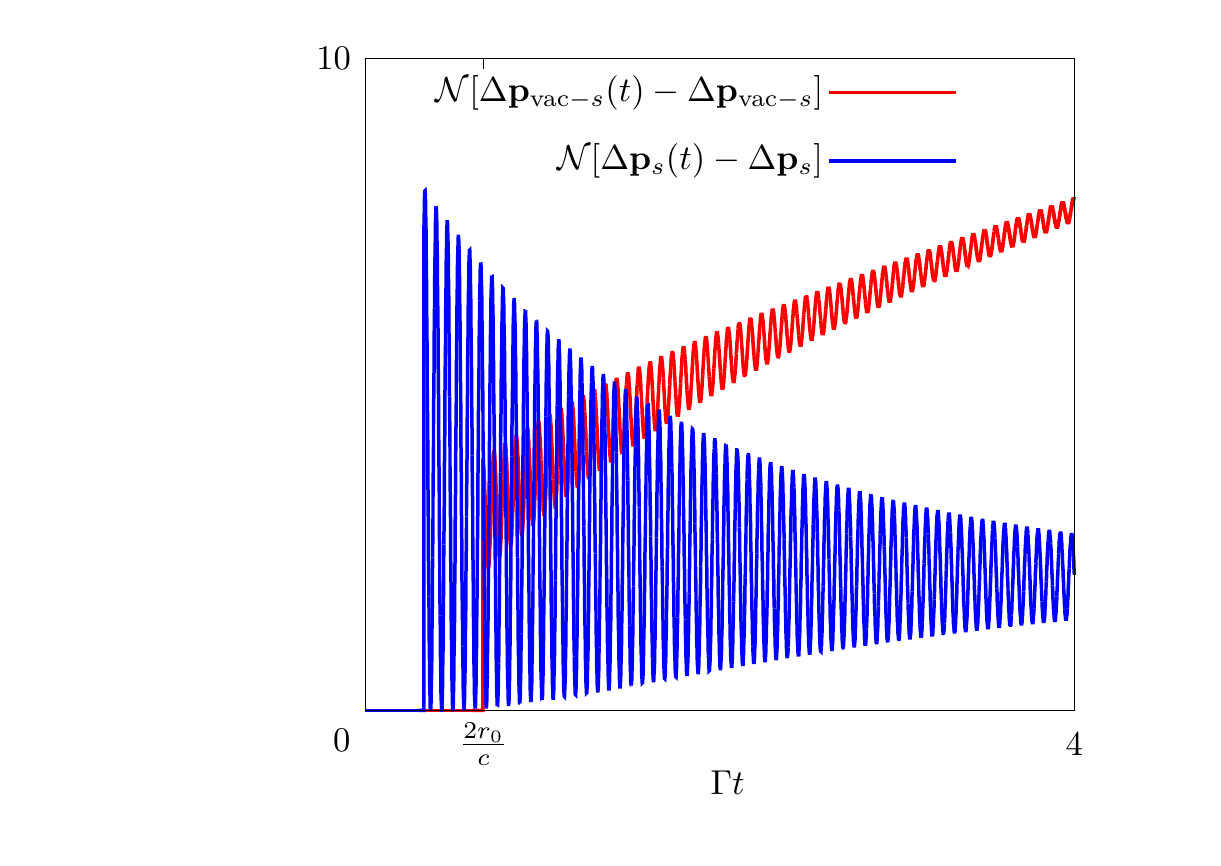} 
\caption{The contributions ${\cal N}[\Delta{\bf p}_s(t)-\Delta{\bf p}_s]$ and ${\cal N}[\Delta{\bf p}_{{\rm vac}-s}(t)-\Delta{\bf p}_{{\rm vac}-s}]$ are plotted as in figure \ref{f1}, but with the larger transition frequency $\omega_0 = 100\Gamma$. These parameters are well within the Markovian regime $\omega_0\gg \Gamma$. Comparing figures \ref{f1} and \ref{f2} we see that for fixed $\Gamma$ increasing $\omega_0/\Gamma$ results in more rapid oscillations, but the same overall behavior with increasing time.}\label{f2}
\end{center}
\end{minipage}
\end{figure}
\begin{figure}[t]
\begin{minipage}{\columnwidth}
\begin{center}
\hspace*{-2.1cm}\includegraphics[scale=0.85]{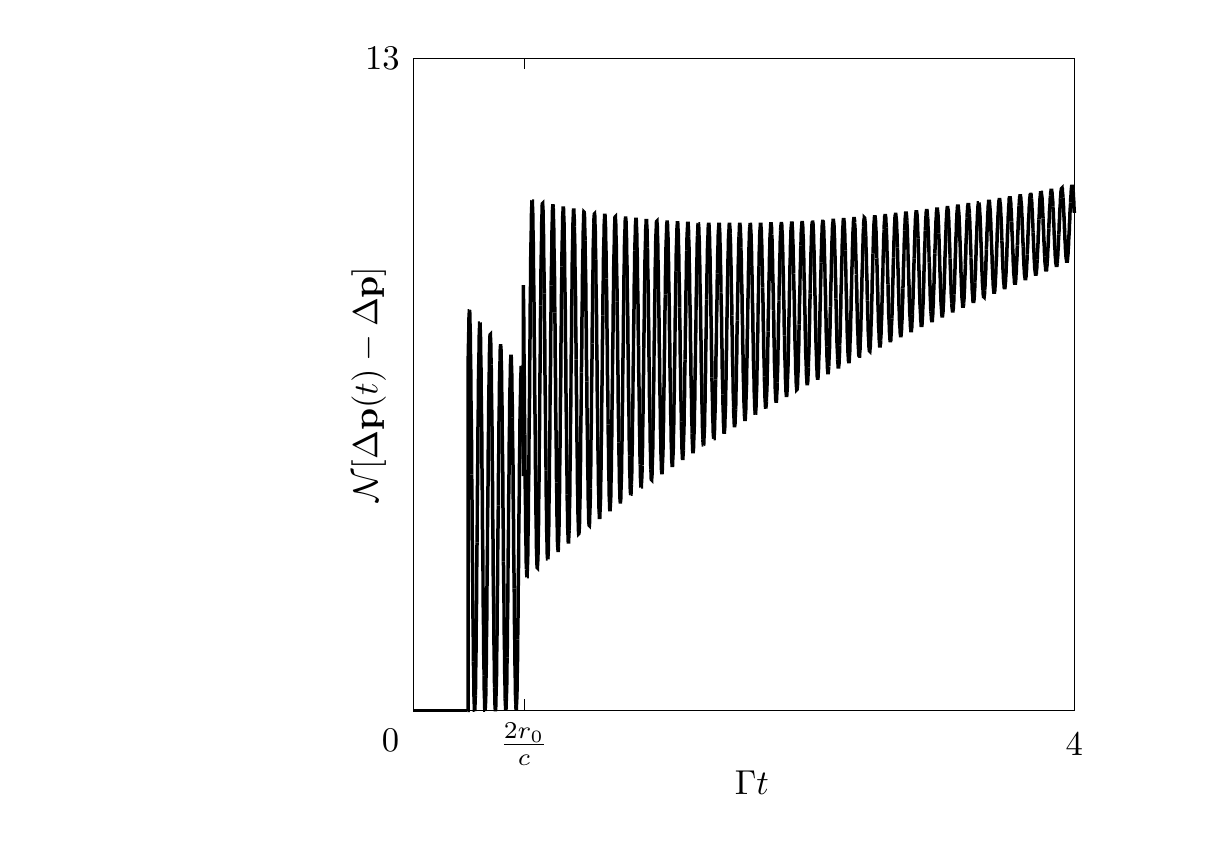} 
\caption{The complete change in momentum dispersion ${\cal N}[\Delta{\bf p}(t)-\Delta{\bf p}]$ is plotted with all parameters chosen as in figure \ref{f2}. Due to the advanced-wave contribution coming from vacuum-source correlations the momentum dispersion increases with time. For large $t$ a good fit for the behaviour of ${\cal N}[\Delta{\bf p}(t)-\Delta{\bf p}]$ is ${\cal N}[\Delta{\bf p}(t)-\Delta{\bf p}]= k+\Gamma t$ where $k$ is a constant.}\label{f3}
\end{center}
\end{minipage}
\end{figure}
%
%
\noindent For $t<2r_0$ the world line of the dipole does not pass through the intersection of the backward lightcone of $(t,{\bf r}_0)$ and the forward lightcone of any point on the charge's world line with time coordinate in the interval $[0,t]$. In this case there are no advanced contributions to $C_{E_iE_i}(t,{\bf r}_0|t',{\bf r}_0)$. 

The change in momentum dispersion is
\begin{align}
\Delta {\bf p}(t) -\Delta{\bf p}= \int_0^t dt' {d \over dt'} \Delta {\bf p}(t'),
\end{align}
which can be partitioned in the same way as the momentum diffusion as the sum of two terms $\Delta{\bf p}_s(t)-\Delta{\bf p}_s$ and $\Delta{\bf p}_{{\rm vac}-s}(t)-\Delta{\bf p}_{{\rm vac}-s}$. These two contributions are shown separately in figures \ref{f1} and \ref{f2}. Both contributions have an oscillatory character, which is increasingly suppressed as time increases.

It is instructive to compare this scenario with that of a free particle whose initial Gaussian wave-packet spreads in time, but whose momentum dispersion is conserved. We see that since the particle is not free the momentum dispersion is oscillatory and is not conserved. However, if we retain only the contribution $\Delta{\bf p}_s(t)-\Delta{\bf p}_s$ that comes from $G_{E_iE_i}$ then in the long time limit when the dipole has completely decayed the charge's momentum dispersion reaches a steady value, effectively behaving like that of a free particle possessing non-zero average momentum. Of course, even if initially saturated the uncertainty principle for a free charge is highly unsaturated for times on the order of $t_q$ due to the spreading of the initial wave-packet.

If the contribution $\Delta{\bf p}_{{\rm vac}-s}(t)-\Delta{\bf p}_{{\rm vac}-s}$ coming from $\Delta_{E_iE_i}$ is also included we see that the behaviour of the momentum dispersion is quite different. In contrast to the case of a free particle the momentum dispersion increases linearly in time in the long-time limit, despite the decay of the initially excited dipole. The change in the dispersion of the position, which quantifies the spreading of the wave-packet is found via integration of Eq.~ (\ref{eext}). Assuming the initial dipole-field state $\ket{e,0}$, it is given by
\begin{align}\label{posdisp}
&\Delta  {\bf r}(t)-\Delta {\bf r}= {t^2\over 2m^2}\Delta {\bf p} +{q^2\over m^2} \nonumber \\ &\times \int_0^t dt_1 \int_0^t dt_2 \int_0^{t_1} dt_3 \int_0^{t_2} dt_4 \, C_{E_iE_i}(t_3,{\bf r}_0|t_4,{\bf r}_0)\nonumber \\ &~+{\rm c.c.}
\end{align}
where we have assumed that there are no initial correlations between the position and momentum of the charge, and where we have again neglected the pure vacuum contribution. As well as the free component $t^2\Delta {\bf p}/m^2$ that results in the usual spreading of the wave packet, there is also a dipole-dependent contribution in Eq.~ (\ref{posdisp}), which oscillates in time. In the radiation zone $r_0\gg \omega_0^{-1}$ the free component dominates, because the remaining contribution decreases as $r_0^{-2}$. As in the case of a free charge the product $\Delta {\bf r}(t) \Delta {\bf p}(t)$ increases with time such that the uncertainty principle becomes increasingly unsaturated.

\section{Conclusions}\label{conc}

In this paper we have focused on quadratic functionals of the electromagnetic field associated with a single stationary dipole. We have extended previous perturbative results, which demonstrate that vacuum-source-field correlations provide significant contributions to the radiation intensity of the dipole. We have derived general non-perturbative expressions for arbitrary quadratic field functionals using standard optical approximations. Due to correlations between the vacuum-field and the source-fields of the dipole, contributions coming from the advanced green's function for the wave-operator are generally non-vanishing within unequal-time field correlation functions. This lies in marked contrast to the derivation of the source-fields themselves.

The contribution of vacuum-source correlations to photo-detection amplitudes was shown to be insignificant. However, by developing a description of a free charge $q$ analogous to the multipolar description of a dipole, it was shown that vacuum-source correlations yield significant contributions to statistical predictions involving the force experienced by a free charge in the field of a dipole.

In classical electrodynamics, while advanced solutions to the wave equation can be used as a theoretical tool, advanced waves are not usually thought to posses any basis in physical reality. Our results indicate that advanced waves associated with the quantum vacuum do exist, and that it should in principle be possible to verify this using an experiment. These results offer yet another signature of the quantum nature of the vacuum and open up interesting prospects for further investigation of advanced-wave like correlations in various branches of quantum optics.

{\em Acknowledgement}. This work was supported by the UK engineering and physical sciences research council grant number EP/N008154/1. I thank Dr. A. Nazir for useful discussions relating to this work.

\onecolumngrid
\vspace{\columnsep}

\section{Appendix}

\subsection{Derivation of source-fields}\label{ap0}

The equation of motion for the photon annihilation operator is found using the Hamiltonian (\ref{h}) and once formally integrated reads
\begin{align}\label{a}
a_\lambda(t,{\bf k}) = a_\lambda({\bf k})e^{-i\omega t} + \int_0^t dt' \, e^{-i\omega(t-t')}\sqrt{\omega\over 2(2\pi)^3}\sum_{nm}{\bf e}_\lambda({\bf k})\cdot {\bf d}_{nm}\sigma_{nm}(t')\equiv a_{\lambda,0}(t,{\bf k})+a_{\lambda,s}(t,{\bf k}).
\end{align}
where $\sigma_{nm}(t) = e^{iHt}\ket{n}\bra{m}e^{-iHt}$. To go further we express Eq.~(\ref{a}) in terms of interaction picture dipole operators ${\tilde \sigma}_{nm}(t) = \sigma_{nm}(t)e^{-i\omega_{nm}t}$ and perform a rotating-wave approximation, which neglects terms oscillating rapidly at frequencies $\omega+\omega_{nm}$, $n>m$. Substituting the resulting expression into Eq.~(\ref{E}), and evaluating the angular integral and polarisation summation yields
\begin{align}\label{er1}
E_{{\rm rad},s,i}^{(+)}(t,{\bf x}) = {i\over 4\pi^2}\sum_{n<m}  \int_0^\infty d\omega \, \omega^3 \tau_{ij}(\omega x)d_{nm}^j \int_0^t dt' e^{-i(\omega+\omega_{nm})(t-t')}{\tilde \sigma}_{nm}(t')e^{i\omega_{nm}t}
\end{align}
where
\begin{align}\label{tau}
\tau_{ij}(\omega x) = (\delta_{ij}-{\hat x}_i{\hat x}_j){\sin (\omega x) \over \omega x} + (\delta_{ij}-3{\hat x}_i{\hat x}_j)\left[{\cos(\omega x) \over (\omega x)^2}-{\sin (\omega x) \over (\omega x)^3}\right]
\end{align}
The near, intermediate and far zone components of expression (\ref{er1}) are evaluated separately. Using the identity $\sin x =(e^{ix}-e^{-ix})/2i$ the far zone component can be written
\begin{align}\label{er2}
{\bf E}_{{\rm rad},s}^{(+)}(t,{\bf x}) = {1\over 8\pi^2 x} \sum_{n<m} [{\bf d}_{nm}-{\hat {\bf x}}({\hat {\bf x}}\cdot {\bf d}_{nm})]\int_0^t dt' \,{\tilde \sigma}_{nm}(t') \int_0^\infty d\omega\, \omega^2 e^{i(\omega+\omega_{nm})t'}\left[e^{-i\omega t_r}- e^{-i\omega t_a}\right]
\end{align}
where $t_r=t-x$ and $t_a=t+x$. This expression involves both retarded and advanced waves but, it will be seen that the advanced waves do not contribute. The integrand in Eq.~(\ref{er2}) is dominated by the resonant contribution at $\omega=\omega_{mn}>0$. We can therefore make use of the following Markov approximation
\begin{align}\label{ma}
\int_0^\infty d\omega \, f(\omega) e^{i\omega t'} \left[e^{-i\omega t_r}\pm e^{-i\omega t_a}\right] &\approx f(\omega_{mn}) \int_{-\infty}^\infty d\omega \, e^{i\omega t'}\left[e^{-i\omega t_r}\pm e^{-i\omega t_a}\right] \nonumber \\ &= 2\pi f(\omega_{mn}) [\delta(t'-t_r)\pm\delta(t'-t_a)],
\end{align}
where $f$ is a suitably behaved function, to obtain the radiation source-field
\begin{align}\label{sc2}
{\bf E}_{{\rm rad},s}(t,{\bf x}) = {\bf E}_{{\rm rad},s}^{(+)}(t,{\bf x})+{\bf E}_{{\rm rad},s}^{(-)}(t,{\bf x}),\qquad  {\bf E}_{{\rm rad},s}^{(+)}(t,{\bf x})={1\over 4\pi x} \sum_{n<m} \omega_{nm}^2 [{\bf d}_{nm}-{\hat {\bf x}}({\hat {\bf x}}\cdot {\bf d}_{nm})]\sigma_{nm}(t_r).
\end{align}
The derivations of the near and intermediate-zone components of the electric source-field, and the magnetic source-field  do not involve any essentially different steps to those above. The final result including both full source-fields is given by Eq.~(\ref{Epms}).

\subsection{Contribution of vacuum source-field correlations to radiated power}\label{ap2}

\subsubsection{Perturbative calculation of vacuum-source-field correlations and associated radiated power}

The equation of motion for the operator $\sigma_{nm}(t)$ can be integrated to give
\begin{align}\label{sigtil}
{\tilde \sigma}_{nm}(t)= \sigma_{nm} + i\int_0^t dt' \, \sum_p\left\{ [{\bf d}_{mp}\cdot {\bf D}_{\rm T}(t',{\bf 0})]{\tilde \sigma}_{np}(t')e^{i\omega_{mp}t'} - [{\bf d}_{pn}\cdot {\bf D}_{\rm T}(t',{\bf 0})]{\tilde \sigma_{pm}}(t')e^{i\omega_{pn}t'} \right\}.
\end{align} 
where
\begin{align}
{\bf D}_{\rm T}(t,{\bf 0}) = i\int d^3 k \sum_\lambda \sqrt{\omega\over 2(2\pi)^3} {\bf e}_\lambda({\bf k}) \left[a_\lambda(t,{\bf k}) - a^\dagger_\lambda(t,{\bf k})\right].
\end{align}
The first order component of Eq.~(\ref{sigtil}) is
\begin{align}\label{sigtil1}
{\tilde \sigma}_{nm}^{(1)}(t) &= i\int_0^t dt' \, \sum_p\left\{ [{\bf d}_{mp}\cdot {\bf D}_{{\rm T},0}(t',{\bf 0})] \sigma_{np}e^{i\omega_{mp}t'} - [{\bf d}_{pn}\cdot {\bf D}_{{\rm T},0}(t',{\bf 0})] \sigma_{pm}e^{i\omega_{pn}t'} \right\}
\end{align}
where
\begin{align}
{\bf D}_{{\rm T},0}(t,{\bf 0}) = i\int d^3 k \sum_\lambda \sqrt{\omega\over 2(2\pi)^3} {\bf e}_\lambda({\bf k}) \left[a_\lambda({\bf k})e^{-i\omega t} - a^\dagger_\lambda({\bf k})e^{i\omega t}\right].
\end{align}
Using Eqs.~(\ref{sc2}) and (\ref{sigtil1}) the radiation source-field correct to second order can now be completely expressed in terms of operators at $t=0$;
\begin{align}\label{e2}
{\bf E}_{{\rm rad},s}^{(2)}(t,{\bf x}) =  {1\over 4\pi x} \sum_{n,m} \omega_{nm}^2 [{\bf d}_{nm}-{\hat {\bf x}}({\hat {\bf x}}\cdot {\bf d}_{nm})]{\tilde \sigma}_{nm}^{(1)}(t_r)e^{i\omega_{nm}t_r}.
\end{align}
Meanwhile the vacuum electric field is
\begin{align}\label{evac}
{\bf E}_0(t,{\bf x}) &= i\int d^3k \sum_\lambda \sqrt{\omega \over 2(2\pi)^3}{\bf e}_\lambda({\bf k})a_\lambda({\bf k})e^{-i\omega t+i{\bf k}\cdot {\bf x}} + {\rm H.c.}.
\end{align}
Using Eqs.~(\ref{e2}) and (\ref{evac}) we obtain
\begin{align}
\langle {\bf E}_0(t,{\bf x})\cdot {\bf E}_{{\rm rad},s}^{(2)}(t,{\bf x})\rangle_{0;e} = {i \over (8\pi^2)^2 x} & \int d^3 k\, \omega\, e^{i{\bf k}\cdot {\bf x}} \sum_n \omega_{en}^2 [{\bf d}_{en}\cdot {\bf e}_\lambda({\bf k})]\, {\bf e}_\lambda({\bf k})\cdot [{\bf d}_{ne}-{\hat {\bf x}}({\hat {\bf x}}\cdot {\bf d}_{ne})]\nonumber \\ &\times \int_0^{t_r} dt' e^{i\omega(t'-t)} \left[e^{i\omega_{en}(t'-t_r)}-e^{-i\omega_{en}(t'-t_r)}\right].
\end{align}
We now evaluate the angular integral and sum over polarisations, and we retain only terms which vary as $x^{-2}$, to give
\begin{align}
\langle {\bf E}_0(t,{\bf x})\cdot {\bf E}_{{\rm rad},s}^{(2)}(t,{\bf x})\rangle_{0;e} =-{1\over 4(2\pi)^3 x^2} \int_0^{t_r} dt' &\sum_n \omega_{en}^2[{\bf d}_{en}-{\hat {\bf x}}({\hat {\bf x}}\cdot {\bf d}_{en})]^2 \left(e^{i\omega_{en}(t'-t_r)}-e^{-i\omega_{en}(t'-t_r)}\right)  \nonumber \\ & \int_0^\infty \,d\omega \, \omega^2\left(e^{i\omega(t'-t_r)}-e^{i\omega(t'-t_a)}\right).
\end{align}
Next we perform a rotating wave-approximation as in the derivation of Eq.~(\ref{er2}), which yields
\begin{align}
\langle {\bf E}_0(t,{\bf x})\cdot {\bf E}_{{\rm rad},s}^{(2)}(t,{\bf x})\rangle_{0;e}&=-{1\over 4(2\pi)^3 x^2} \int_0^{t_r} dt' \int_0^\infty \,d\omega \, \omega^2\left(e^{i\omega(t'-t_r)}-e^{i\omega(t'-t_a)}\right) \nonumber \\ &\times \left[\sum_{n>e} \omega_{en}^2[{\bf d}_{en}-{\hat {\bf x}}({\hat {\bf x}}\cdot {\bf d}_{en})]^2 e^{i\omega_{en}(t'-t_r)}-\sum_{n<e} \omega_{en}^2[{\bf d}_{en}-{\hat {\bf x}}({\hat {\bf x}}\cdot {\bf d}_{en})]^2 e^{-i\omega_{en}(t'-t_r)}\right].
\end{align}
Since the integrand is now dominated by positive resonant frequencies $\omega=\omega_{en},~n<e$ and $\omega=-\omega_{en},~n>e$, we can perform the Markov approximation, Eq.~(\ref{ma}), to obtain
\begin{align}\label{pvsf}
\langle {\bf E}_0(t,{\bf x})\cdot {\bf E}_{{\rm rad},s}^{(2)}(t,{\bf x})\rangle_{0;e}={1\over 2} {1\over (4\pi x)^2} \sum_n {\rm sgn}(\omega_{en})\omega^4_{en}[{\bf d}_{en}-{\hat {\bf x}}({\hat {\bf x}}\cdot {\bf d}_{en})]^2
\end{align}
where we have used
\begin{align}\label{delta}
\int_0^t dt' \delta(t'-t)f(t')={1\over 2}f(t).
\end{align}
From Eq.~(\ref{pvsf}) we obtain the corresponding contribution to the radiated power;
\begin{align}
P_{\rm vac-s} = \int d\Omega \, x^2\left(\langle {\bf E}_0(t,{\bf x})\cdot {\bf E}_{{\rm rad},s}^{(2)}(t,{\bf x})\rangle_{0;e} +{\rm c.c.}\right) = {1\over 2}\sum_n {\rm sgn}(\omega_{en})\omega_{en}\Gamma_{en}.
\end{align}
which is Eq.~(\ref{Pvac}).

\subsubsection{Radiation reaction: derivation of the Heisenberg-Langevin equations}

Restricting our attention to a two-level dipole with ground state $\ket{g}$ and excited state $\ket{e}$, we define the dipole operators
\begin{align}
\sigma^+ = \ket{e}\bra{g},\qquad \sigma^- \ket{g}\bra{e},\qquad \sigma^z = [\sigma^+,\sigma^-].
\end{align}
The two-level transition frequency, dipole moment and decay rate are denoted $\omega_0$, ${\bf d}$ and $\Gamma$ respectively. From the solution (\ref{a}) we calculate the dipole's own reaction field as
\begin{align}
{\bf D}^{(+)}_{{\rm T},s}(t,{\bf 0}) &= i\int d^3 k \sqrt{\omega\over 2(2\pi)^3}\sum_\lambda {\bf e}_{\lambda}({\bf k}) a_{\lambda,s}(t,{\bf k}) \nonumber \\ &= i{\Gamma  \over 2\pi} { {\hat {\bf d}} \over d}\int_0^\infty d\omega\, \left({\omega\over \omega_0}\right)^3\int_0^t dt'\, e^{i\omega(t'-t)}\left[{\tilde \sigma}^+(t')e^{i\omega_0 t'}+{\tilde \sigma}^-(t')e^{-i\omega_0 t'}\right].
\end{align}
where we have evaluated the angular integral and polarisation summation. We now perform the rotating-wave approximation and then the Markov approximation, Eq.~(\ref{ma}), which give
\begin{align}\label{ast}
{\bf D}^{(+)}_{{\rm T},s}(t,{\bf 0}) = i\int d^3 k \sqrt{\omega\over 2(2\pi)^3}\sum_\lambda {\bf e}_\lambda({\bf k})  a_{\lambda,s}(t,{\bf k}) = i{\Gamma \over 2}{ {\hat {\bf d}} \over d} \sigma^-(t)
\end{align}
where we have used used Eq.~(\ref{delta}). Substituting Eq.~(\ref{ast}) into the Heisenberg equation for each of $\sigma^+$ and $\sigma^z$ yields the following equations of motion
\begin{align}
{\dot \sigma}^+(t) &= \left(i\omega_0-{\Gamma\over 2}\right)\sigma^+(t) + {\Gamma\over 2}\sigma^-(t) + \sigma_{\rm vac}'^+(t)\label{prwa} \\ 
{\dot \sigma^z}(t) &= -\Gamma[\sigma^z(t)+1] + \sigma'^z_{\rm vac}(t)\label{prwa2}
\end{align}
where
\begin{align}
&\sigma'^+_{\rm vac}(t) = \int d^3k\sum_\lambda \sqrt{\omega\over 2(2\pi)^2} {\bf d}\cdot {\bf e}_\lambda({\bf k})\left[a_\lambda^\dagger({\bf k})\sigma^z(t)e^{i\omega t} -\sigma^z(t) a_\lambda({\bf k})e^{-i\omega t}\right] \\ 
&\sigma'^z_{\rm vac}(t) = \int d^3k\sum_\lambda \sqrt{2\omega\over (2\pi)^2} {\bf d}\cdot {\bf e}_\lambda({\bf k})\left[a_\lambda^\dagger({\bf k})[\sigma^+(t)-\sigma^-(t)]e^{i\omega t} -[\sigma^+(t)-\sigma^-(t)] a_\lambda({\bf k})e^{-i\omega t}\right].
\end{align}
Making the rotating-wave approximation in Eqs.~(\ref{prwa}) and (\ref{prwa2}) yields optical Heisenberg-Langevin type equations (with no external driving). Formally integrating these equations then yields
\begin{align}
\sigma^+(t) = e^{(i\omega_0-\Gamma/2)t}\sigma^+ +\int_0^t dt' e^{(i \omega_0-\Gamma/2)(t-t')}\sigma^+_{\rm vac}(t')\label{sol1b} \\ 
\sigma^z(t) +1 = e^{-\Gamma t}[\sigma^z +1]+ \int_0^t dt' e^{-\Gamma (t-t')}\sigma^z_{\rm vac}(t').\label{sol2b}
\end{align}
where
\begin{align}\label{sigzvacb}
&\sigma^+_{\rm vac}(t) = \int d^3k\sum_\lambda \sqrt{\omega\over 2(2\pi)^2} {\bf d}\cdot {\bf e}_\lambda({\bf k})a_\lambda^\dagger({\bf k})\sigma^z(t)e^{i\omega t} =i[{\bf d}\cdot {\bf D}^{(-)}_{{\rm T},0}(t,{\bf 0})]\sigma^z(t) \\ 
&\sigma^z_{\rm vac}(t) = -\int d^3k\sum_\lambda \sqrt{2\omega\over (2\pi)^2} {\bf d}\cdot {\bf e}_\lambda({\bf k})\left[a_\lambda^\dagger({\bf k})\sigma^-(t)e^{i\omega t}+\sigma^+(t) a_\lambda({\bf k})e^{-i\omega t}\right] = -2i[{\bf d}\cdot {\bf D}^{(-)}_{{\rm T},0}(t,{\bf 0})]\sigma^-(t) + {\rm H.c.}
\end{align}
Making the rotating-wave approximation yields the solutions in Eqs~~(\ref{sol1}) and (\ref{sol2}).

\subsubsection{Non-perturbative calculation of vacuum-source-field correlations and associated radiated power}

We now use the solutions (\ref{sol1}) and (\ref{sol2}) to calculate the radiated power. We note that although Markov and rotating wave approximations have been employed we have not used perturbation theory. The pure source-field contribution to the intensity is immediately found using the solution (\ref{sc2}) restricted to two dipole levels;
\begin{align}
\langle {\bf E}_{{\rm rad},s}(t,{\bf x})\cdot {\bf E}_{{\rm rad},s}(t,{\bf x}) \rangle_{0,e} &= \left({\omega^2_0 \over 4\pi x }[{\bf d} - {\hat {\bf x}}({\bf d}\cdot {\hat {\bf x}})]\right)^2 \langle \sigma^+(t_r)\sigma^-(t_r) +\sigma^-(t_r)\sigma^+(t_r)\rangle_{0;e} \nonumber \\ &= \left({\omega^2_0 \over 4\pi x }[{\bf d} - {\hat {\bf x}}({\bf d}\cdot {\hat {\bf x}})]\right)^2.
\end{align}
The corresponding contribution to the power is then easily obtained as $P_s = \omega_0\Gamma/2$. The vacuum source-field correlation function is found using Eqs.~(\ref{evac}) and (\ref{sc2}) to be
\begin{align}\label{co}
\langle {\bf E}_0(t,{\bf x})\cdot {\bf E}_{{\rm rad},s}(t,{\bf x}) \rangle_{0,e} &=\langle {\bf E}^{(+)}_0(t,{\bf x})\cdot {\bf E}_{{\rm rad},s}(t,{\bf x}) \rangle_{0,e} = {\omega_0^2 \over 4\pi x}[{\bf d} - {\hat {\bf x}}({\bf d}\cdot {\hat {\bf x}})]\cdot \langle {\bf E}^{(+)}_0(t,{\bf x})[\sigma^+(t_r)+\sigma^-(t_r)]\rangle_{0;e}.
\end{align}
From Eqs.~(\ref{sol1}) and (\ref{sigzvac}) it follows that
\begin{align}
\langle {\bf E}^{(+)}_0(t,{\bf x}&)[\sigma^+(t_r)+\sigma^-(t_r)]\rangle_{0;e} \nonumber \\ &= \langle {\bf E}^{(+)}_0(t,{\bf x})\sigma^+(t_r)\rangle_{0;e} =i\int d^3 k \sqrt{\omega \over 2(2\pi)^3} \sum_\lambda {\bf e}_\lambda({\bf k}) \langle a_\lambda({\bf k})\sigma^+(t_r)\rangle_{0;e}e^{-i\omega t+i{\bf k}\cdot{\bf x}} \nonumber \\ &= i\int d^3 k \sum_\lambda {\omega \over 2(2\pi)^3} {\bf e}_\lambda({\bf k}) [{\bf e}_\lambda({\bf k}) \cdot {\bf d}]e^{i{\bf k}\cdot {\bf x}} \int_0^{t_r} dt'\,\langle \sigma^z(t') \rangle_{0;e}\, e^{-i\omega(t- t')}e^{(i \omega_0-\Gamma/2)(t_r-t')}
\end{align}
Substituting this result into Eq.~(\ref{co}) we obtain
\begin{align}\label{B}
&\langle {\bf E}^{(+)}_0(t,{\bf x})\cdot {\bf E}_{{\rm rad},s}(t,{\bf x}) \rangle_{0,e}=\langle {\bf E}^{(+)}_0(t,{\bf x})\cdot {\bf E}^{(-)}_{{\rm rad},s}(t,{\bf x}) \rangle_{0,e} \nonumber \\ &= {i\omega_0^2 \over 4\pi x} \int d^3 k \sum_\lambda {\omega \over 2(2\pi)^3} ({\bf e}_\lambda({\bf k})\cdot {\bf d}) ({\bf e}_\lambda({\bf k}) \cdot [{\bf d} - {\hat {\bf x}}({\bf d}\cdot {\hat {\bf x}})])e^{i{\bf k}\cdot {\bf x}}\int_0^{t_r} dt' (2e^{-\Gamma t'}-1)e^{(i \omega_0-\Gamma/2)(t_r-t')}e^{-i\omega(t-t')}
\end{align}
where we have used $\langle \sigma^z(t)\rangle_{0;e} = 2e^{-\Gamma t}-1$, which follows from Eq.~(\ref{sol2}). We see from Eq.~(\ref{B}) that the rotating-wave and Markov approximations do not eliminate the anti-normally ordered contribution $\langle {\bf E}^{(+)}_0(t,{\bf x})\cdot {\bf E}^{(-)}_{{\rm rad},s}(t,{\bf x}) \rangle_{0,e}$, which is a slowly varying resonant contribution. It is nonetheless not included in the correlation function $\langle {\bf E}^{(-)}(t,{\bf x})\cdot {\bf E}^{(+)}(t,{\bf x}) \rangle_{0,e}$. Evaluating the angular integral and polarisation summation now yields
\begin{align}
\langle& {\bf E}^{(+)}_0(t,{\bf x})\cdot {\bf E}^{(-)}_{{\rm rad},s}(t,{\bf x}) \rangle_{0,e} \nonumber \\ &= \left({\omega_0 \over 4\pi x }[{\bf d} - {\hat {\bf x}}({\bf d}\cdot {\hat {\bf x}})]\right)^2 &\int_0^{t_r} dt' \left(2e^{-\Gamma t'}-1\right)e^{(i\omega_0-\Gamma/2)(t_r-t')} {1\over 2\pi}\int_0^\infty d\omega \, \omega^2\left[e^{-i\omega(t_r-t')}-e^{-i\omega(t_a-t')}\right].
\end{align}
We now make use of the Markov approximation, Eq.~(\ref{ma}), which gives
\begin{align}\label{3}
\langle {\bf E}^{(+)}_0(t,{\bf x})\cdot {\bf E}^{(-)}_{{\rm rad},s}(t,{\bf x}) \rangle_{0,e} &= \left({\omega_0^2 \over 4\pi x }[{\bf d} - {\hat {\bf x}}({\bf d}\cdot {\hat {\bf x}})]\right)^2 \left(e^{-\Gamma t_r}-{1\over 2}\right)
\end{align}
where we have used Eq.~(\ref{delta}). From Eq.~(\ref{3}) we obtain the corresponding contribution to the radiated power;
\begin{align}
P_{\rm vac-s} = \int d\Omega \, x^2\left(\langle {\bf E}^{(+)}_0(t,{\bf x})\cdot {\bf E}^{(-)}_{{\rm rad},s}(t,{\bf x})\rangle_{0;e} +{\rm c.c.}\right) = \omega_0 \Gamma \left(e^{-\Gamma t_r}- {1\over 2}\right),
\end{align}
which is Eq.~(\ref{pvs}).

\subsection{Unequal-time commutators of electromagnetic fields and advanced-wave correlations}\label{ap3}

\subsubsection{Unequal-time commutators}

We assume that dipole operators $\sigma^\pm(t)$ and $\sigma^z(t)$ commute with the field operators $a_\lambda(t,{\bf k})$ and $a_\lambda^\dagger(t,{\bf k})$. We demonstrate here how various commutation relations between the fields ${\bf E}^\pm, {\bf B}^\pm$ at arbitrary space-time points can be calculated. As our example we take the electric field commutator $[E_i^{(+)}(t,{\bf x}),E^{(-)}_j(t',{\bf x}')]$, which can be written as the sum of four terms;
\begin{align}\label{com4}
&[E_i^{(+)}(t,{\bf x}),E^{(-)}_j(t',{\bf x}')] \nonumber \\ &=[E_{0,i}^{(+)}(t,{\bf x}),E^{(-)}_{0,j}(t',{\bf x}')]+[E_{0,i}^{(+)}(t,{\bf x}),E^{(-)}_{s,j}(t',{\bf x}')]+[E_{s,i}^{(+)}(t,{\bf x}),E^{(-)}_{0,j}(t',{\bf x}')]+[E_{s,i}^{(+)}(t,{\bf x}),E^{(-)}_{s,j}(t',{\bf x}')].
\end{align}
The pure vacuum term is easily found using Eq.~(\ref{Dfree}) to be
\begin{align}\label{c1}
[E_{0,i}^{(+)}(t,{\bf x}),E^{(-)}_{0,j}(t',{\bf x}')]=\int d^3 k \sum_\lambda {\omega \over 2(2\pi)^3} e^{-i\omega(t-t')+i{\bf k}\cdot ({\bf x}-{\bf x}')}. 
\end{align}
The pure source term is also easily found with the help of Eq.~(\ref{Ema}) to be
\begin{align}\label{c2}
[E_{s,i}^{(+)}(t,{\bf x}),E^{(-)}_{s,j}(t',{\bf x}')]={\cal E}_i^*({\bf x}){\cal E}_j({\bf x}')\theta(t_r)\theta(t_r')[\sigma^-(t_r),\sigma^+(t_r')]
\end{align}
where $t_r=t-x$ and $t_r'=t'-x'$.

Simplification of the remaining two terms on the right-hand-side of Eq.~(\ref{com4}), which involve both the vacuum and source-fields requires more work. We begin by noting that by rearranging the rotating-wave approximated integrated Heisenberg equation for the operator $a_\lambda(t,{\bf k})$, we can express $a_\lambda({\bf k})$ as
\begin{align}\label{aat0}
a_\lambda({\bf k}) = e^{i\omega t}a_\lambda(t,{\bf k}) - \int_0^t dt' e^{i\omega t'} \sqrt{\omega \over 2(2\pi)^3} [{\bf e}_\lambda({\bf k})\cdot {\bf d}]\sigma^-(t').
\end{align}
Noting that $E^{(-)}_{s,j}(t',{\bf x}')$ is a function of $\sigma^+(t_r')$ we use Eq.~(\ref{aat0}) to express the vacuum field as a function of $a_\lambda(t_r',{\bf k})$, which allows us to write the term $[E_{0,i}^{(+)}(t,{\bf x}),E^{(-)}_{s,j}(t',{\bf x}')]$ as
\begin{align}
[E_{0,i}^{(+)}(t,{\bf x}),E^{(-)}_{s,j}(t',{\bf x}')] = -i{\cal E}_j({\bf x}')\theta(t_r')\int d^3 k \sum_\lambda {\omega \over 2(2\pi)^3}e_{\lambda,i}({\bf k})[{\bf e}_\lambda({\bf k})\cdot {\bf d}]e^{i{\bf k}\cdot {\bf x}} \int_0^{t_r'} ds\, e^{-i\omega(t-s)}[\sigma^-(s),\sigma^+(t_r')]
\end{align}
where we have used the fact that equal-time dipole and field operators commute. We now evaluate the polarisation summation and integration over solid angle to obtain
\begin{align}
[E_{0,i}^{(+)}(t,{\bf x}),E^{(-)}_{s,j}(t',{\bf x}')] = -{i\over 4\pi^2}{\cal E}_j({\bf x}')d_k \theta(t_r')\int_0^\infty d\omega \, \omega^3 \tau_{ik}(\omega x)\int_0^{t_r'} ds\, e^{-i\omega(t-s)}[\sigma^-(s),\sigma^+(t_r')]
\end{align}
where $\tau_{ik}$ is defined in Eq.~(\ref{tau}). Using Eq.~(\ref{tau}) yields
\begin{align}
[E_{0,i}^{(+)}&(t,{\bf x}),E^{(-)}_{s,j}(t',{\bf x}')] \nonumber \\ &= -{1\over 8\pi^2}{\cal E}_j({\bf x}')d_k \theta(t_r')\int_0^\infty d\omega \, \int_0^{t_r'} ds\, [\sigma^-(s),\sigma^+(t_r')]  \bigg[{\omega^2\over x}(\delta_{ik}-{\hat x}_i{\hat x}_k)\left(e^{i\omega(s-t_r)}-e^{i\omega(s-t_a)}\right)\nonumber \\ &+{i\omega\over x^2}(\delta_{ik}-3{\hat x}_i{\hat x}_k)\left(e^{i\omega(s-t_r)}+e^{i\omega(s-t_a)}\right)  -{1\over x^3}(\delta_{ik}-3{\hat x}_i{\hat x}_k)\left(e^{i\omega(s-t_r)}-e^{i\omega(s-t_a)}\right)\bigg],
\end{align}
and using the Markov approximation we obtain
\begin{align}\label{c3}
[E_{0,i}^{(+)}(t,{\bf x})&,E^{(-)}_{s,j}(t',{\bf x}')] \nonumber \\ \approx& -{1\over 4\pi}{\cal E}_j({\bf x}')d_k \theta(t_r') \int_0^{t_r'} ds\, [\sigma^-(s),\sigma^+(t_r')] \bigg[{\omega^2_0\over x}(\delta_{ik}-{\hat x}_i{\hat x}_k)\left[\delta(s-t_r)-\delta(s-t_a)\right]\nonumber \\ &+{i\omega_0\over x^2}(\delta_{ik}-3{\hat x}_i{\hat x}_k)\left[\delta(s-t_r)+\delta(s-t_a)\right]  -{1\over x^3}(\delta_{ik}-3{\hat x}_i{\hat x}_k)\left[\delta(s-t_r)-\delta(s-t_a)\right]\bigg] \nonumber \\ =& -{1\over 4\pi}{\cal E}_j({\bf x}')d_k\theta(t_r') \theta(t_r)\theta(t_r'-t_r)[\sigma^-(t_r),\sigma^+(t_r')]\left[{\omega^2_0\over x}(\delta_{ik}-{\hat x}_i{\hat x}_k)+\left({i\omega_0\over x^2}-{1\over x^3}\right)(\delta_{ik}-3{\hat x}_i{\hat x}_k)\right] \nonumber \\ & -{1\over 4\pi}{\cal E}_j({\bf x}')d_k \theta(t_a)\theta(t_r'-t_a)[\sigma^-(t_a),\sigma^+(t_r')]\left[-{\omega^2_0\over x}(\delta_{ik}-{\hat x}_i{\hat x}_k)+\left({i\omega_0\over x^2}+{1\over x^3}\right)(\delta_{ik}-3{\hat x}_i{\hat x}_k)\right] \nonumber \\ =&-{\cal E}_i^*({\bf x}){\cal E}_j({\bf x}')\theta(t_r')\theta(t_r)\theta(t_r'-t_r)[\sigma^-(t_r),\sigma^+(t_r')] +{\cal E}_i({\bf x}){\cal E}_j({\bf x}')\theta(t_r')\theta(t_a)\theta(t_r'-t_a)[\sigma^-(t_a),\sigma^+(t_r')].
\end{align}
This result coincides with the result given in Eq.~(\ref{retadv1}) with ${\bf X}={\bf E}={\bf Y}$. The remaining term $ [E_{s,i}^{(+)}(t,{\bf x}),E^{(-)}_{0,j}(t',{\bf x}')]$ on the right-hand-side of Eq.~(\ref{com4}) can be calculated in the same way as the term calculated above and is given by the right-hand-side of Eq.~(\ref{retadv2}) with ${\bf X}={\bf E}={\bf Y}$. Thus, having calculated all source-dependent terms on the right-hand-side of Eq.~(\ref{com4}) we have using Eqs.~(\ref{c2}) and (\ref{c3}) that
\begin{align}\label{DEE}
&\Delta_{E_iE_j}(t,{\bf x}|t',{\bf x}') = [E_i^{(+)}(t,{\bf x}),E^{(-)}_j(t',{\bf x}')] -[E_{0,i}^{(+)}(t,{\bf x}),E^{(-)}_{0,j}(t',{\bf x}')] \nonumber \\ &= {\cal E}_i({\bf x}){\cal E}_j({\bf x}')\theta(t_r')\theta(t_a)\theta(t_r'-t_a)[\sigma^-(t_a),\sigma^+(t_r')] +{\cal E}^*_i({\bf x}){\cal E}^*_j({\bf x}')\theta(t_r)\theta(t'_a)\theta(t_r-t'_a)[\sigma^-(t_r),\sigma^+(t_a')],
\end{align}
which coincides with Eq.~(\ref{DXYfin}) in the case ${\bf X}={\bf E}={\bf Y}$. Calculations of the same type as above can be used to find the remaining unequal time commutators $[X_i^{(+)}(t,{\bf x}),Y_j^{(-)}(t',{\bf x}')],~{\bf X},{\bf Y}={\bf E},{\bf B}$, which eventually yields Eqs.~(\ref{retadv1}), (\ref{retadv2}) and (\ref{DXYfin}). By setting $t=t'$ in Eq.~(\ref{DXYfin}) the right-hand-side vanishes, which proves that $[X_i^{(+)}(t,{\bf x}),Y^{(-)}_j(t,{\bf x}')] =[X_{0,i}^{(+)}(t,{\bf x}),Y^{(-)}_{0,j}(t,{\bf x}')]$ within the Markov and rotating-wave approximations. Thus, the approximate theory is formally consistent as claimed at the end of section \ref{eandbfs}.

\subsubsection{Advanced-wave correlations}

Supposing that the initial state of the dipole and field is $\ket{0,e}$ the average $\langle \Delta_{X_iY_j}(t,{\bf x}|t',{\bf x}')\rangle_{e,0}$ depends, according to Eq.~(\ref{DXYfin}), on the averages $\langle [\sigma^-(t_a),\sigma^+(t_r')]\rangle_{e,0}$ and $\langle [\sigma^-(t_r),\sigma^+(t_a')]\rangle_{e,0}$. These averages can be found using Eqs.~(\ref{sol1}) and (\ref{sol2}). The averages $\langle\sigma^+(t_r')\sigma^-(t_a)\rangle_{e,0}$ and $\langle\sigma^+(t_a')\sigma^-(t_r)\rangle_{e,0}$ are immediately found to be
\begin{align}\label{d0}
\langle\sigma^+(t_r')\sigma^-(t_a)\rangle_{e,0} = e^{(i\omega_0 -\Gamma/2)t_r'}e^{(-i\omega_0 -\Gamma/2)t_a},\qquad \langle\sigma^+(t_a')\sigma^-(t_r)\rangle_{e,0} = e^{(i\omega_0 -\Gamma/2)t_a}e^{(-i\omega_0 -\Gamma/2)t_r'}.
\end{align}
The average $\langle \sigma^-(t_a)\sigma^+(t_r')\rangle_{e,0}$ is
\begin{align}\label{d1}
\langle \sigma^-(t_a)\sigma^+(t_r')\rangle_{e,0} = d_i d_j \int_0^{t_a} ds \int_0^{t_r'} ds' e^{(-i\omega_0-\Gamma/2)(t_a-s)}e^{(i\omega_0-\Gamma/2)(t_r'-s')}\langle \sigma^z(s) D^{(+)}_{{\rm T},0,i}(s,{\bf 0})D^{(-)}_{{\rm T},0,j}(s',{\bf 0})\sigma^z(s')\rangle_{e,0}.
\end{align}
Using the commutator $[D^{(+)}_{{\rm T},0,i}(s,{\bf 0}),D^{(-)}_{{\rm T},0,j}(s',{\bf 0})]$ Eq.~(\ref{d1}) can be written as the sum of two terms;
\begin{align}\label{d2}
\langle \sigma^-&(t_a)\sigma^+(t_r')\rangle_{e,0} \nonumber \\ &= \int_0^{t_a} ds \int_0^{t_r'} ds'\int d^3 k \sum_\lambda {\omega \over 2(2\pi)^3}[{\bf e}_{\lambda}({\bf k})\cdot {\bf d}]^2 e^{(-i\omega_0-\Gamma/2)(t_a-s)}e^{(i\omega_0-\Gamma/2)(t_r'-s')}e^{-i\omega(s-s')} \langle \sigma^z(s)\sigma^z(s')\rangle_{e,0} \nonumber \\ &~~~+\,  d_i d_j \int_0^{t_a} ds \int_0^{t_r'} ds' e^{(-i\omega_0-\Gamma/2)(t_a-s)}e^{(i\omega_0-\Gamma/2)(t_r'-s')}\langle \sigma^z(s) D^{(-)}_{{\rm T},0,j}(s',{\bf 0}) D^{(+)}_{{\rm T},0,i}(s,{\bf 0})\sigma^z(s')\rangle_{e,0}.
\end{align}
Performing the angular integration and polarisation summation and employing the Markov approximation in the first term on the right-hand-side yields
\begin{align}\label{d3}
\Gamma \int_0^{t_a} ds \int_0^{t_r'} ds' \delta(s-s')e^{(-i\omega_0-\Gamma/2)(t_a-s)}e^{(i\omega_0-\Gamma/2)(t_r'-s')}\langle \sigma^z(s)\sigma^z(s')\rangle_{e,0}.
\end{align}
Since within $\Delta_{X_iY_j}(t,{\bf x}|t',{\bf x}')$ this term is multiplied by $\theta(t_r'-t_a)$ we can assume that $t_r'\geq t_a$ in the above. Performing the $s$ integral then yields
\begin{align}\label{d4}
\Gamma e^{(-i\omega_0-\Gamma/2)t_a}e^{(i\omega_0-\Gamma/2)t_r'} \int_0^{t_a} ds' e^{\Gamma s'} =  e^{(-i\omega_0-\Gamma/2)t_a}e^{(i\omega_0-\Gamma/2)t_r'}(e^{\Gamma t_a}-1).
\end{align}

To show that the second term on the right-hand-side of Eq.~(\ref{d2}) vanishes one can show that for $s\neq s'$ it is always possible to change the order of $\sigma^z(s) D^{(-)}_{{\rm T},0,j}(s',{\bf 0})$ or of $D^{(+)}_{{\rm T},0,i}(s,{\bf 0})\sigma^z(s')$ within the expectation value. The integral in Eq.~(\ref{d2}), which involves this expectation value must therefore vanish. First consider the commutator $[\sigma^z(s), D^{(-)}_{{\rm T},0,j}(s',{\bf 0})]$. Using Eq.~(\ref{aat0}) we can express this as
\begin{align}
[\sigma^z(s), D^{(-)}_{{\rm T},0,j}(s',{\bf 0})]=i\int d^3 k \sum_\lambda  {\omega \over 2(2\pi)^3} e_{\lambda,j}({\bf k}) [{\bf e}_\lambda({\bf k})\cdot {\bf d}] \int_0^s ds''[\sigma^z(s),\sigma^+(s'')]e^{i\omega(s'-s'')}
\end{align}
Writing this expression in terms of the interaction picture operator ${\tilde \sigma}^+(s'')$ allows us to use the Markov approximation, which gives
\begin{align}\label{com1}
[\sigma^z(s), D^{(-)}_{{\rm T},0,j}(s',{\bf 0})]=i{{\hat d}_j \over d}\Gamma \int_0^s ds'' \delta(s''-s')[\sigma^z(s),\sigma^+(s'')] = i{{\hat d}_j \over d}\Gamma [\sigma^z(s),\sigma^+(s')]\theta(s-s')\theta(s').
\end{align}
Similarly the commutator $[D^{(+)}_{{\rm T},0,i}(s,{\bf 0}),\sigma^z(s')]$ is found to be
\begin{align}\label{com2}
i{{\hat d}_i \over d}\Gamma [\sigma^+(s),\sigma^z(s')]\theta(s'-s)\theta(s).
\end{align}
Since $\langle \sigma^z(s) D^{(-)}_{{\rm T},0,j}(s',{\bf 0}) D^{(+)}_{{\rm T},0,i}(s,{\bf 0})\sigma^z(s')\rangle_{e,0}=\langle [\sigma^z(s) ,D^{(-)}_{{\rm T},0,j}(s',{\bf 0}) ][D^{(+)}_{{\rm T},0,i}(s,{\bf 0}),\sigma^z(s')]\rangle_{e,0}$ using Eqs.~(\ref{com1}) and (\ref{com2}) the second term on the right-hand-side of Eq.~(\ref{d2}) can be expressed in the form
\begin{align}
\int_0^{t_a} ds \int_0^{t_r'} ds' f(s,s')\theta(s-s')\theta(s'-s) = 0.
\end{align}
Thus, the second term on the right-hand-side of Eq.~(\ref{d2}) vanishes within the Markov approximation. As a result $\langle \sigma^-(t_a)\sigma^+(t_r')\rangle_{e,0}$ is given by the right-hand-side of Eq.~(\ref{d4}). Using Eq.~(\ref{d0}) and (\ref{d4}) we therefore have
\begin{align}
\langle [\sigma^-(t_a),\sigma^+(t_r')]\rangle_{e,0}=e^{(-i\omega_0-\Gamma/2)t_a}e^{(i\omega_0-\Gamma/2)t_r'}(e^{\Gamma t_a}-2)
\end{align}
in agreement with Eq.~(\ref{DXYexp}). The remaining commutator $\langle [\sigma^-(t_r),\sigma^+(t_a')]\rangle_{e,0}$ can be found in a similar fashion and this yields the remaining part of Eq.~(\ref{DXYexp}).

\subsection{Static precursors within Coulomb gauge electric source-field of a charge}\label{ap4}

The Hamiltonian (\ref{hq}) yields the following solution for the Coulomb gauge operator $a_\lambda(t,{\bf k})$
\begin{align}\label{ac}
a_\lambda(t,{\bf k}) = e^{-i\omega t}a_\lambda ({\bf k}) + i\int_0^t dt' e^{-i\omega(t-t')} {1\over \sqrt{2\omega}}{\bf e}_\lambda({\bf k}) \cdot {\tilde {\bf J}}(t',{\bf k})
\end{align}
where ${\tilde {\bf J}}$ denotes the Fourier transform of the current ${\bf J}$;
\begin{align}
{\tilde {\bf J}}(t,{\bf k}) = {1\over \sqrt{(2\pi)^3}} \int d^3 x\, {\bf J}(t,{\bf x})e^{-i{\bf k}\cdot {\bf x}},\qquad {\bf J}(t,{\bf x}) = {q\over 2}\left[{\dot {\bf r}}(t)\delta({\bf x}-{\bf r}(t)) + \delta({\bf x}-{\bf r}(t)){\dot {\bf r}(t)}\right].
\end{align}
When substituted into the mode-expansion for the transverse electric field Eq.~(\ref{ac}) yields the source-field
\begin{align}
{\bf E}_{{\rm T},s}(t,{\bf x}) = - \int d^3 k \int_0^t dt' {1\over 2\sqrt{(2\pi)^3}}\left({\tilde {\bf J}}(t',{\bf k})-{\hat {\bf k}}[{\hat {\bf k}}\cdot {\tilde {\bf J}}(t',{\bf k})]\right)e^{i{\bf k}\cdot {\bf x}}e^{-i\omega(t-t')} + {\rm H.c.}
\end{align}
Using integration by parts and the continuity equation ${\dot \rho}=-\nabla \cdot {\bf J}$ where $\rho(t,{\bf x}) = q\delta({\bf x}-{\bf r}(t))$ is the charge density, this can be written
\begin{align}\label{etrans}
&{\bf E}_{{\rm T},s}(t,{\bf x}) = \left[{i\over \sqrt{(2\pi)^3}}\int d^3 k\, {{\hat {\bf k}}{\tilde \rho}(t',{\bf k})\over \omega}\cos[\omega(t-t')] e^{i{\bf k}\cdot {\bf x}}\right]_{t'=0}^{t'=t} \nonumber \\ &-{1\over \sqrt{(2\pi)^3}}\int_0^t dt' {d\over dt}\int d^3 k \, {\tilde {\bf J}}(t',{\bf k}){\sin[\omega(t-t')] \over \omega}e^{i{\bf k}\cdot {\bf x}} - {\nabla\over \sqrt{(2\pi)^3}}\int d^3 k \int_0^t dt' {\tilde \rho}(t',{\bf k}){\sin[\omega(t-t')] \over \omega}e^{i{\bf k}\cdot {\bf x}}.
\end{align}
The function $\sin[\omega(t-t')]/\sqrt{(2\pi)^3}\omega$ is essentially the Fourier transform of the green's function for the wave-operator. Using the convolution theorem and Gauss' law ${\tilde {\bf E}}_{\rm L}(t,{\bf k}) = -i{\hat {\bf k}}{\tilde \rho}(t,{\bf k})/\omega$ Eq.~(\ref{etrans}) can be written
\begin{align}\label{etrans2}
{\bf E}_{{\rm T},s}(t,{\bf x}) =&-{\bf E}_{\rm L}(t,{\bf x}) +{d\over dt}\int d^3 x' {\bf E}_{\rm L}(0,{\bf x}')G(t,{\bf x}|0,{\bf x}') \nonumber \\ &-{d\over dt}\int d^3 x' \int_0^t dt' {\bf J}(t',{\bf x}')G^+(t,{\bf x}|t',{\bf x}')-\nabla\int d^3 x' \int_0^t dt' \rho(t',{\bf x}')G^+(t,{\bf x}|t',{\bf x}')
\end{align}
where $G=G^++G^-$, and where we have used
\begin{align}\label{pre}
-{i\over \sqrt{(2\pi)^3}}&\int d^3 k\, {{\hat {\bf k}}{\tilde \rho}(0,{\bf k})\over \omega}\cos[\omega t] e^{i{\bf k}\cdot {\bf x}}={1\over \sqrt{(2\pi)^3}}\int d^3 k\,{\tilde {\bf E}}_{\rm L}(0,{\bf k})\cos[\omega t] e^{i{\bf k}\cdot {\bf x}}\nonumber \\ &={d\over dt}{1\over \sqrt{(2\pi)^3}}\int d^3 k\,{\tilde {\bf E}}_{\rm L}(0,{\bf k}){\sin[\omega t]\over \omega} e^{i{\bf k}\cdot {\bf x}} = {d\over dt}\int d^3 x' {\bf E}_{\rm L}(0,{\bf x}')G(t,{\bf x}|0,{\bf x}').
\end{align}
By adding the longitudinal field to ${\bf E}_{{\rm T},s}(t,{\bf x})$ we obtain the total electric source-field ${\bf E}_s(t,{\bf x})$. Along with the well-known retarded source-field given by the second line in Eq.~(\ref{etrans2}), the electric source-field also possesses a term given in Eq.~(\ref{pre}), which is dependent on the charge density at the initial time $t=0$. 

\vspace{\columnsep}

\bibliography{advanced.bib}

\begin{thebibliography}{36}
\providecommand{\natexlab}[1]{#1}
\providecommand{\url}[1]{\texttt{#1}}
\expandafter\ifx\csname urlstyle\endcsname\relax
  \providecommand{\doi}[1]{doi: #1}\else
  \providecommand{\doi}{doi: \begingroup \urlstyle{rm}\Url}\fi

\bibitem[Barton(1989)]{barton_elements_1989}
G.~Barton.
\newblock \emph{Elements of {Green}'s {Functions} and {Propagation}:
  {Potentials}, {Diffusion}, {And} {Waves}}.
\newblock Oxford University Press, Oxford : New York, new ed edition edition,
  July 1989.
\newblock ISBN 978-0-19-851998-0.

\bibitem[Baxter et~al.(1990)Baxter, Babiker, and Loudon]{baxter_gauge_1990}
C.~Baxter, M.~Babiker, and R.~Loudon.
\newblock Gauge {Invariant} {QED} with {Arbitrary} {Mixing} of p.a and q.e
  {Interactions}.
\newblock \emph{Journal of Modern Optics}, 37\penalty0 (4):\penalty0 685--699,
  1990.
\newblock ISSN 0950-0340.
\newblock \doi{10.1080/09500349014550761}.
\newblock URL
  \url{http://www.tandfonline.com/doi/abs/10.1080/09500349014550761}.

\bibitem[Bialynicki-Birula(1996)]{bialynicki-birula_v_1996}
Iwo Bialynicki-Birula.
\newblock V {Photon} {Wave} {Function}.
\newblock In E.~Wolf, editor, \emph{Progress in {Optics}}, volume~36, pages
  245--294. Elsevier, January 1996.
\newblock URL
  \url{http://www.sciencedirect.com/science/article/pii/S0079663808703160}.
\newblock DOI: 10.1016/S0079-6638(08)70316-0.

\bibitem[Carmichael(1999)]{carmichael_statistical_1999}
Howard~J. Carmichael.
\newblock \emph{Statistical {Methods} in {Quantum} {Optics} 1}.
\newblock Springer Berlin Heidelberg, Berlin, Heidelberg, 1999.
\newblock ISBN 978-3-642-08133-0 978-3-662-03875-8.
\newblock URL \url{http://link.springer.com/10.1007/978-3-662-03875-8}.
\newblock DOI: 10.1007/978-3-662-03875-8.

\bibitem[Casimir and Polder(1948)]{casimir_influence_1948}
H.~B.~G. Casimir and D.~Polder.
\newblock The {Influence} of {Retardation} on the {London}-van der {Waals}
  {Forces}.
\newblock \emph{Physical Review}, 73\penalty0 (4):\penalty0 360--372, February
  1948.
\newblock \doi{10.1103/PhysRev.73.360}.
\newblock URL \url{https://link.aps.org/doi/10.1103/PhysRev.73.360}.

\bibitem[Cohen-Tannoudji et~al.(1997)Cohen-Tannoudji, Dupont-Roc, and
  Grynberg]{cohen-tannoudji_photons_1997}
Claude Cohen-Tannoudji, Jacques Dupont-Roc, and Gilbert Grynberg.
\newblock \emph{Photons and atoms: introduction to quantum electrodynamics}.
\newblock Wiley VCH, March 1997.
\newblock ISBN 0-471-18433-0.

\bibitem[Dalibard et~al.(1982)Dalibard, Dupont-Roc, and
  Cohen-Tannoudji]{dalibard_vacuum_1982}
J.~Dalibard, J.~Dupont-Roc, and C.~Cohen-Tannoudji.
\newblock Vacuum fluctuations and radiation reaction : identification of their
  respective contributions.
\newblock \emph{Journal de Physique}, 43\penalty0 (11):\penalty0 1617--1638,
  November 1982.
\newblock ISSN 0302-0738.
\newblock \doi{10.1051/jphys:0198200430110161700}.
\newblock URL \url{http://dx.doi.org/10.1051/jphys:0198200430110161700}.

\bibitem[Davies(1975)]{davies_scalar_1975}
P.~C.~W. Davies.
\newblock Scalar production in {Schwarzschild} and {Rindler} metrics.
\newblock \emph{Journal of Physics A: Mathematical and General}, 8\penalty0
  (4):\penalty0 609, 1975.
\newblock ISSN 0305-4470.
\newblock \doi{10.1088/0305-4470/8/4/022}.
\newblock URL \url{http://stacks.iop.org/0305-4470/8/i=4/a=022}.

\bibitem[Drummond(1987)]{drummond_unifying_1987}
P.~D. Drummond.
\newblock Unifying the p.a and q.e interactions in photodetector theory.
\newblock \emph{Physical Review A}, 35\penalty0 (10):\penalty0 4253--4262, May
  1987.
\newblock \doi{10.1103/PhysRevA.35.4253}.
\newblock URL \url{http://link.aps.org/doi/10.1103/PhysRevA.35.4253}.

\bibitem[Glauber(1963)]{glauber_quantum_1963}
Roy~J. Glauber.
\newblock The {Quantum} {Theory} of {Optical} {Coherence}.
\newblock \emph{Physical Review}, 130\penalty0 (6):\penalty0 2529--2539, June
  1963.
\newblock \doi{10.1103/PhysRev.130.2529}.
\newblock URL \url{http://link.aps.org/doi/10.1103/PhysRev.130.2529}.

\bibitem[Hawking(1975)]{hawking_particle_1975}
S.~W. Hawking.
\newblock Particle creation by black holes.
\newblock \emph{Communications in Mathematical Physics}, 43\penalty0
  (3):\penalty0 199--220, August 1975.
\newblock ISSN 0010-3616, 1432-0916.
\newblock \doi{10.1007/BF02345020}.
\newblock URL \url{https://link.springer.com/article/10.1007/BF02345020}.

\bibitem[Jackson(1998)]{jackson_classical_1998}
John~David Jackson.
\newblock \emph{Classical electrodynamics}.
\newblock Wiley, 3 edition, August 1998.
\newblock ISBN 0-471-30932-X.

\bibitem[Jaffe(2005)]{jaffe_casimir_2005}
R.~L. Jaffe.
\newblock Casimir effect and the quantum vacuum.
\newblock \emph{Physical Review D}, 72\penalty0 (2):\penalty0 021301, July
  2005.
\newblock \doi{10.1103/PhysRevD.72.021301}.
\newblock URL \url{https://link.aps.org/doi/10.1103/PhysRevD.72.021301}.

\bibitem[Mandel(1958)]{mandel_fluctuations_1958}
L.~Mandel.
\newblock Fluctuations of {Photon} {Beams} and their {Correlations}.
\newblock \emph{Proceedings of the Physical Society}, 72\penalty0 (6):\penalty0
  1037, 1958.
\newblock ISSN 0370-1328.
\newblock \doi{10.1088/0370-1328/72/6/312}.
\newblock URL \url{http://stacks.iop.org/0370-1328/72/i=6/a=312}.

\bibitem[Mandel et~al.(1964)Mandel, Sudarshan, and Wolf]{mandel_theory_1964}
L.~Mandel, E.~C.~G. Sudarshan, and E.~Wolf.
\newblock Theory of photoelectric detection of light fluctuations.
\newblock \emph{Proceedings of the Physical Society}, 84\penalty0 (3):\penalty0
  435, 1964.
\newblock ISSN 0370-1328.
\newblock \doi{10.1088/0370-1328/84/3/313}.
\newblock URL \url{http://stacks.iop.org/0370-1328/84/i=3/a=313}.

\bibitem[Milonni et~al.(1995)Milonni, James, and
  Fearn]{milonni_photodetection_1995}
P.~W. Milonni, D.~F.~V. James, and H.~Fearn.
\newblock Photodetection and causality in quantum optics.
\newblock \emph{Physical Review A}, 52\penalty0 (2):\penalty0 1525--1537,
  August 1995.
\newblock \doi{10.1103/PhysRevA.52.1525}.
\newblock URL \url{http://link.aps.org/doi/10.1103/PhysRevA.52.1525}.

\bibitem[Milonni(1994)]{milonni_quantum_1994}
Peter~W Milonni.
\newblock \emph{The quantum vacuum: an introduction to quantum
  electrodynamics}.
\newblock Academic Press, Boston, 1994.
\newblock ISBN 0-12-498080-5 978-0-12-498080-8.

\bibitem[Milonni et~al.(1973)Milonni, Ackerhalt, and
  Smith]{milonni_interpretation_1973}
Peter~W. Milonni, Jay~R. Ackerhalt, and Wallace~Arden Smith.
\newblock Interpretation of {Radiative} {Corrections} in {Spontaneous}
  {Emission}.
\newblock \emph{Physical Review Letters}, 31\penalty0 (15):\penalty0 958--960,
  October 1973.
\newblock \doi{10.1103/PhysRevLett.31.958}.
\newblock URL \url{http://link.aps.org/doi/10.1103/PhysRevLett.31.958}.

\bibitem[Moskalenko et~al.(2015)Moskalenko, Riek, Seletskiy, Burkard, and
  Leitenstorfer]{moskalenko_paraxial_2015}
A. S. Moskalenko, C.~Riek, D. V. Seletskiy, G.~Burkard, and
  A.~Leitenstorfer.
\newblock Paraxial {Theory} of {Direct} {Electro}-optic {Sampling} of the
  {Quantum} {Vacuum}.
\newblock \emph{Physical Review Letters}, 115\penalty0 (26):\penalty0 263601,
  December 2015.
\newblock \doi{10.1103/PhysRevLett.115.263601}.
\newblock URL \url{https://link.aps.org/doi/10.1103/PhysRevLett.115.263601}.

\bibitem[Olson and Ralph(2012)]{olson_extraction_2012}
S.~Jay Olson and Timothy~C. Ralph.
\newblock Extraction of timelike entanglement from the quantum vacuum.
\newblock \emph{Physical Review A}, 85\penalty0 (1):\penalty0 012306, January
  2012.
\newblock \doi{10.1103/PhysRevA.85.012306}.
\newblock URL \url{https://link.aps.org/doi/10.1103/PhysRevA.85.012306}.

\bibitem[Power and Thirunamachandran(1992)]{power_quantum_1992}
E.~A. Power and T.~Thirunamachandran.
\newblock Quantum electrodynamics with nonrelativistic sources. {IV}.
  {Poynting} vector, energy densities, and other quadratic operators of the
  electromagnetic field.
\newblock \emph{Physical Review A}, 45\penalty0 (1):\penalty0 54--63, January
  1992.
\newblock \doi{10.1103/PhysRevA.45.54}.
\newblock URL \url{http://link.aps.org/doi/10.1103/PhysRevA.45.54}.

\bibitem[Power and Thirunamachandran(1993)]{power_quantum_1993}
E.~A. Power and T.~Thirunamachandran.
\newblock Quantum electrodynamics with nonrelativistic sources. {V}.
  {Electromagnetic} field correlations and intermolecular interactions between
  molecules in either ground or excited states.
\newblock \emph{Physical Review A}, 47\penalty0 (4):\penalty0 2539--2551, April
  1993.
\newblock \doi{10.1103/PhysRevA.47.2539}.
\newblock URL \url{http://link.aps.org/doi/10.1103/PhysRevA.47.2539}.

\bibitem[Power and Thirunamachandran(1999)]{power_time_1999}
E.~A. Power and T.~Thirunamachandran.
\newblock Time dependence of operators in minimal and multipolar
  nonrelativistic quantum electrodynamics. {I}. {Electromagnetic} fields in the
  neighborhood of an atom.
\newblock \emph{Physical Review A}, 60\penalty0 (6):\penalty0 4927--4935,
  December 1999.
\newblock \doi{10.1103/PhysRevA.60.4927}.
\newblock URL \url{http://link.aps.org/doi/10.1103/PhysRevA.60.4927}.

\bibitem[Power(1966)]{power_zero-point_1966}
Edwin~A. Power.
\newblock Zero-{Point} {Energy} and the {Lamb} {Shift}.
\newblock \emph{American Journal of Physics}, 34\penalty0 (6):\penalty0
  516--518, 1966.
\newblock \doi{10.1119/1.1973082}.
\newblock URL \url{http://link.aip.org/link/?AJP/34/516/1}.

\bibitem[Reznik et~al.(2005)Reznik, Retzker, and Silman]{reznik_violating_2005}
Benni Reznik, Alex Retzker, and Jonathan Silman.
\newblock Violating {Bell}'s inequalities in vacuum.
\newblock \emph{Physical Review A}, 71\penalty0 (4):\penalty0 042104, April
  2005.
\newblock \doi{10.1103/PhysRevA.71.042104}.
\newblock URL \url{https://link.aps.org/doi/10.1103/PhysRevA.71.042104}.

\bibitem[Riek et~al.(2015)Riek, Seletskiy, Moskalenko, Schmidt, Krauspe,
  Eckart, Eggert, Burkard, and Leitenstorfer]{riek_direct_2015}
C.~Riek, D.~V. Seletskiy, A.~S. Moskalenko, J.~F. Schmidt, P.~Krauspe,
  S.~Eckart, S.~Eggert, G.~Burkard, and A.~Leitenstorfer.
\newblock Direct sampling of electric-field vacuum fluctuations.
\newblock \emph{Science}, 350\penalty0 (6259):\penalty0 420--423, October 2015.
\newblock ISSN 0036-8075, 1095-9203.
\newblock \doi{10.1126/science.aac9788}.
\newblock URL \url{http://science.sciencemag.org/content/350/6259/420}.

\bibitem[Riek et~al.(2017)Riek, Sulzer, Seeger, Moskalenko, Burkard, Seletskiy,
  and Leitenstorfer]{riek_subcycle_2017}
C.~Riek, P.~Sulzer, M.~Seeger, A.~S. Moskalenko, G.~Burkard, D.~V. Seletskiy,
  and A.~Leitenstorfer.
\newblock Subcycle quantum electrodynamics.
\newblock \emph{Nature}, 541\penalty0 (7637):\penalty0 376--379, January 2017.
\newblock ISSN 0028-0836.
\newblock \doi{10.1038/nature21024}.
\newblock URL
  \url{http://www.nature.com/nature/journal/v541/n7637/full/nature21024.html}.

\bibitem[Salam(2008)]{salam_molecular_2008}
A.~Salam.
\newblock Molecular quantum electrodynamics in the {Heisenberg} picture: a
  field theoretic viewpoint.
\newblock \emph{International Reviews in Physical Chemistry}, 27\penalty0
  (3):\penalty0 405--448, 2008.
\newblock ISSN 0144-235X.
\newblock \doi{10.1080/01442350802045206}.
\newblock URL
  \url{http://www.tandfonline.com/doi/abs/10.1080/01442350802045206}.

\bibitem[Salam(2009)]{salam_molecular_2009}
Akbar Salam.
\newblock \emph{Molecular {Quantum} {Electrodynamics}: {Long}-{Range}
  {Intermolecular} {Interactions}}.
\newblock Wiley, 1 edition, November 2009.
\newblock ISBN 0-470-25930-2.

\bibitem[Sipe(1995)]{sipe_photon_1995}
J.~E. Sipe.
\newblock Photon wave functions.
\newblock \emph{Physical Review A}, 52\penalty0 (3):\penalty0 1875--1883,
  September 1995.
\newblock \doi{10.1103/PhysRevA.52.1875}.
\newblock URL \url{https://link.aps.org/doi/10.1103/PhysRevA.52.1875}.

\bibitem[Spohn(2007)]{spohn_dynamics_2007}
Herbert Spohn.
\newblock \emph{Dynamics of charged particles and their radiation field}.
\newblock Cambridge University Press, June 2007.
\newblock ISBN 0-521-03707-7.

\bibitem[Stokes(2016)]{stokes_quantum_2016}
Adam Stokes.
\newblock Quantum optical dipole radiation fields.
\newblock \emph{European Journal of Physics}, 37\penalty0 (3):\penalty0 034001,
  2016.
\newblock ISSN 0143-0807.
\newblock \doi{10.1088/0143-0807/37/3/034001}.
\newblock URL \url{http://stacks.iop.org/0143-0807/37/i=3/a=034001}.

\bibitem[Stokes et~al.(2012)Stokes, Kurcz, Spiller, and
  Beige]{stokes_extending_2012}
Adam Stokes, Andreas Kurcz, Tim~P. Spiller, and Almut Beige.
\newblock Extending the validity range of quantum optical master equations.
\newblock \emph{Physical Review A}, 85\penalty0 (5):\penalty0 053805, May 2012.
\newblock \doi{10.1103/PhysRevA.85.053805}.
\newblock URL \url{http://link.aps.org/doi/10.1103/PhysRevA.85.053805}.

\bibitem[Unruh(1976)]{unruh_notes_1976}
W.~G. Unruh.
\newblock Notes on black-hole evaporation.
\newblock \emph{Physical Review D}, 14\penalty0 (4):\penalty0 870--892, August
  1976.
\newblock \doi{10.1103/PhysRevD.14.870}.
\newblock URL \url{https://link.aps.org/doi/10.1103/PhysRevD.14.870}.

\bibitem[Welton(1948)]{welton_observable_1948}
Theodore~A. Welton.
\newblock Some {Observable} {Effects} of the {Quantum}-{Mechanical}
  {Fluctuations} of the {Electromagnetic} {Field}.
\newblock \emph{Physical Review}, 74\penalty0 (9):\penalty0 1157--1167,
  November 1948.
\newblock \doi{10.1103/PhysRev.74.1157}.
\newblock URL \url{http://link.aps.org/doi/10.1103/PhysRev.74.1157}.

\bibitem[Zangwill(2012)]{zangwill_modern_2012}
Andrew Zangwill.
\newblock \emph{Modern {Electrodynamics}}.
\newblock Cambridge University Press, Cambridge, December 2012.
\newblock ISBN 978-0-521-89697-9.

\end{thebibliography}

 \end{document}